\documentclass{ws-ijmpcs}

\newcommand\be{\begin{equation}}
\newcommand\ee{\end{equation}}

\usepackage{graphicx}

\begin{document}

\markboth{Elena Amato}
{The origin of Galactic Cosmic Rays}

%
\catchline{}{}{}{}{}
%

\title{THE ORIGIN OF GALACTIC COSMIC RAYS}

\author{ELENA AMATO}

\address{INAF - Osservatorio Astrofisico di Arcetri, Largo E. Fermi, 5, I-50125, Firenze, Italy\\
amato@arcetri.astro.it}

\maketitle

\begin{history}
\received{}
\revised{}
\end{history}

\begin{abstract}
Initial discovery of CRs dates back to a century ago (1912). Their identification as particles rather than radiation dates to about 20 years later and in 20 more years also the first suggestion that they were associated with SNRs was in place. The basic mechanism behind their acceleration was suggested almost 40 years ago. Much work has been done since then to the aim of proving that both the acceleration mechanism and site are well understood, but no definite proof has been obtained: in spite of impressive progress of both theory and observations, the evidence in support of the commonly accepted interpretation is only circumstantial. In the following I will try to make the point on where we stand in terms of how our theories confront with data: I will review recent progress on the subject and try to point to avenues to pursue in order to gather new proofs, if not smoking gun evidence of the origin of Galactic Cosmic Rays.

\end{abstract}

\ccode{PACS numbers: 98.38.Mz,98.38.Fs,96.50.Pw}

\section{Introduction}
\label{sec:intro}	
At the beginning of last century, an outstanding mystery was associated with the inexplicable rate of discharge of electroscopes, which required some source of ionizing radiation to be explained. In 1912, the Austrian scientist Viktor Hess showed that this radiation was of extraterrestrial origin, by carrying electrometers on balloon flights and finding that the rate of discharge was increasing with height. A similar conclusion had already been reached the year before by the italian scientist Domenico Pacini\cite{pacini}, who had carried electrometers on a submarine, showing that the radiation intensity decreased with increasing depth under the sea. This latter studies remained however poorly known and the paternity of the discovery was attributed to Viktor Hess who was awarded the Nobel Prize for it in 1936. 

The term ``Cosmic Rays'' was then coined by Robert Millikan who believed that the unknown ionizing radiation was made of photons, rather than matter. Millikan kept this belief until his death, even after compelling evidence of the contrary had been gathered, and the Cosmic Rays have kept this name, in spite of not being rays.

In the early '30s an association between Cosmic Rays (CRs hereafter) and Supernova (SN) explosions was proposed for the first time\cite{baade}. The proposal was based on an energetic argument\cite{ginsyro61}: the energy density of CRs ($\sim 3 \times 10^{40} {\rm erg}\ {\rm s}^{-1}$) could easily be supplied by SNe if $10\%$ of the explosion energy were turned into accelerated particles (this is assuming an energy release of $10^{51} {\rm erg}$ per SN event and a rate of $1/100\ yr^{-1}$ in the Galaxy). A quantitative proposal for how this energy conversion would take place had to wait until the late '70s when a number of scientists\cite{krymskii,blandostr,bell78a,bell78b} independently suggested that the process of diffusive shock acceleration (also known as ``$1^{\rm st}$ order Fermi mechanism'') taking place at the blast wave launched by a Supernova explosion could provide the required conversion mechanism.

The idea that Supernova Remnant (SNR) shocks are the primary sites of CR acceleration in the Galaxy is what is generally referred to as "Supernova remnant paradigm for the origin of Cosmic Rays". This paradigm has been under scrutiny now for about 50 years, but only in the last few years some clear evidence in its favour has been found. 

\begin{figure}[h!!!]
\includegraphics[scale=.45]{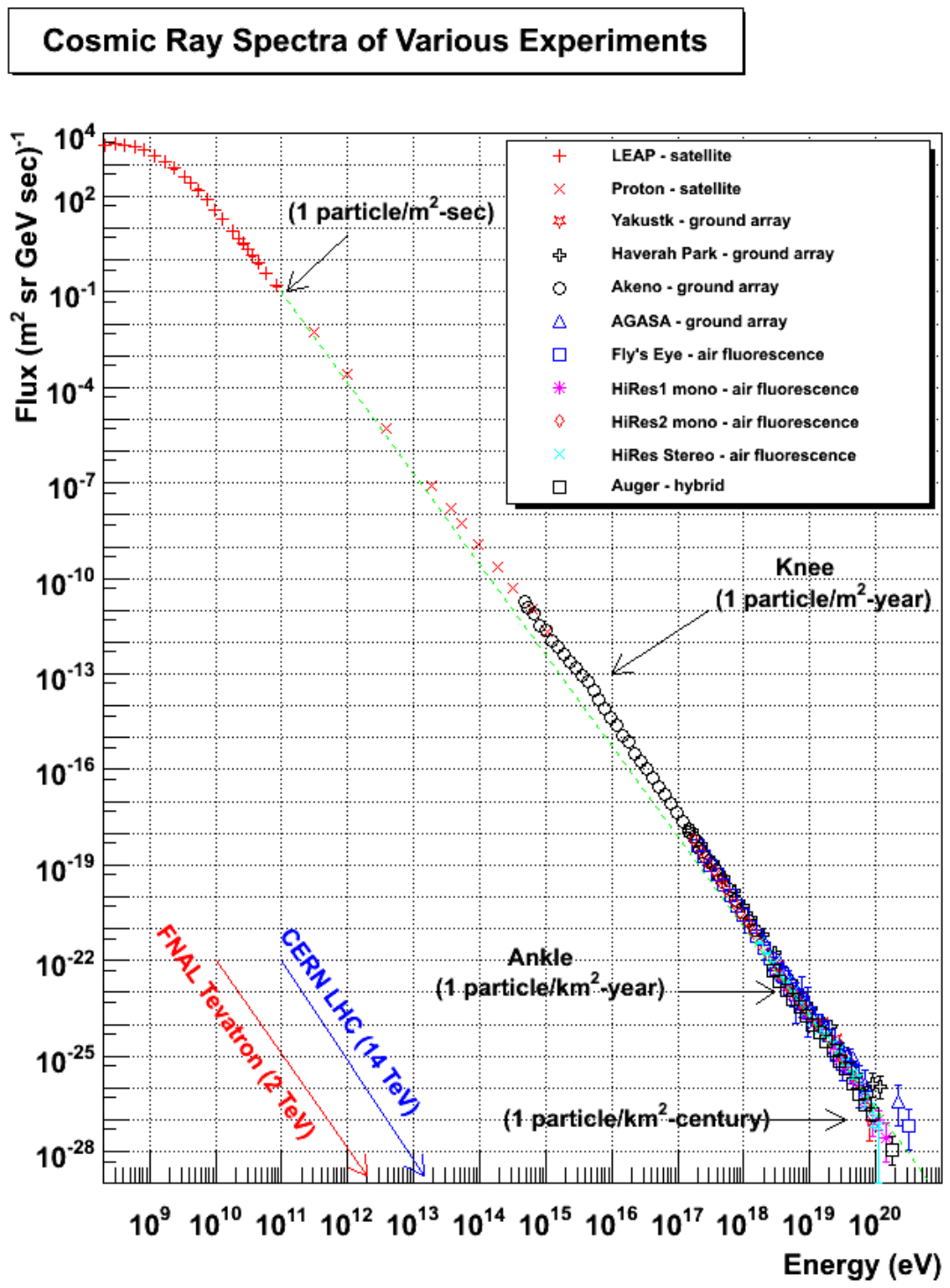}
\caption{Measured spectrum of Cosmic Rays from Ref.~8.}
\label{fig:crspec}
\end{figure}

The CR spectrum, reported in Fig.~\ref{fig:crspec}, has been traditionally described as a single power-law for several decades in energy, between $\sim 30$ GeV and $\sim 5 \times 10^{15}$ eV. Below 30 GeV the spectrum bends as a consequence of solar modulation (the Sun magnetic field partially screens Earth from CRs), while at PeV energies a steepening appears, the so-called {\it knee}. At this energy, the CR energy spectrum (number of particles per unit energy interval), changes from $\propto E^{-2.7}$ to $\propto E^{-3.1}$, and this steepening seems to be accompanied by a change in composition, with the latter becoming heavier. Current data suggest that the {\it knee} can be easily interpreted if the CR acceleration mechanism in our Galaxy is rigidity dependent and the maximum energy protons can reach is $\sim 5 \times 10^{15}$ eV. Then heavier nuclei would reach $Z$ times larger energies, with $Z$ the charge of the nucleus. In this scheme, the heaviest nuclei, namely Fe nuclei, could reach an energy 26 times larger than protons, and the {\it knee} would result as the superposition of the cut-offs of different species.

If the {\it knee} marks the end of the Galactic proton spectrum, then galactic sources must be able to provide acceleration up to that energy. As will be discussed in more detail later on in this review, if SNRs are the main sources of CRs, effective amplification of the interstellar magnetic field must take place. The last generation of X-ray telescopes, with their superb spatial resolution, have not only confirmed the presence in SNRs of electrons with TeV energies, but have also finally allowed to highlight the presence of amplified magnetic fields in these sources, likely associated with efficient acceleration of hadrons\cite{ballet06,vink12}. Also suggestive of efficient acceleration of protons is the measurement of anomalous widths of Balmer lines in some ${\rm H}_\alpha$ bright remnants\cite{helder09,heng10}. Finally, in very recent times, the impressive advances in gamma-ray astronomy have led, for the first time, to direct observational evidence of the presence of mildly relativistic protons in a few SNRs interacting with molecular clouds\cite{tavani10,ackermann13}.
 
This fundamental progress deriving from observations of SNRs has been paralleled by discoveries coming from direct observation of CRs. The latter have touched both the hadronic and leptonic component of CRs. As far as nuclei are concerned, the paradigm of a featureless spectrum at energies below the {\it knee} has been disclaimed by balloon and satellite observations\cite{creamhard,pamelahard} showing a spectral hardening of all species at around 200 GeV/nucleon and a different spectrum for the two most abundant species, protons and He nuclei, with the latter being systematically flatter. These features might be revealing us important clues on the process behind particle acceleration and propagation in the Galaxy.

As for the leptons, a discovery that has been followed by much interest and excitement is that of a rising fraction of positrons versus electrons at energies above a few GeV\cite{pamelapos}. Such a behaviour is not expected unless a source of primary positrons comes into play, and the debate immediately got heated up about the nature of this source, in principle either a class of astrophysical objects or a particle physics process related to dark matter.

At the same time, important progress has been made in terms of the theory that is relevant to explain all the above mentioned phenomena. Comparison between recent theoretical progress and new observations will be the main subject of this review. The literature is vast and the list of references given here is not at all exhaustive. Discussion of other aspects of the subject and more references can be found in other excellent recent reviews, such as Refs.~\protect\refcite{blasirev}\protect and \protect\refcite{zweibelrev}\protect.

In Sec.~\ref{sec:acc} and \ref{sec:nldsa} I will review our current description of CR acceleration at SNR shocks. I will describe the main features of Fermi I mechanism when it operates in conditions of efficient particle acceleration, the so called non-linear diffusive shock acceleration (NLDSA) regime. We will see that an essential feature for SNR shocks to be able to accelerate particles up to the {\it knee} energy is effective amplification of interstellar magnetic fields by a factor of 30-100 (Sec.~\ref{sec:mfastd}) and we will work out the CR spectrum that SNRs are expected to release in the ISM (Sec.~\ref{sec:escape}). In Sec.~\ref{sec:thvsdata} we will see how the basic theoretical predictions compare with observations: the comparison will be successful in many respects but we will also find a few discrepancies in the light of recent $\gamma$-ray observations of SNRs and of new insight on the CR spectra that must be injected in the Galaxy. In Sec.~\ref{sec:nldsarev} I will illustrate how the basic theory can be revised to overcome the discrepancies and in Sec.~\ref{sec:mfanew} I will discuss further revisions that are prompted by new insights in the process of magnetic field amplification. 

In the final part of this review I will discuss some exciting developments prompted by recent observations. In Sec.~\ref{sec:neutrals} I will illustrate the recent progress made in the description of shocks propagating in a partially ionized medium and the potential to gather direct information on CR acceleration from optical observations of Balmer dominated SNRs. Finally, in Sec.~\ref{sec:directcr} I will discuss two major recent discoveries coming from direct observations of Galactic CRs, namely the presence of a hardening in the spectra of protons and He nuclei at an energy of about 200 GeV, and the rise with energy of the fraction of positrons in the leptonic CR component at energies above 30 GeV. I will conclude in Sec.~\ref{sec:concl}.

\section{The acceleration mechanism}
\label{sec:acc}
The most commonly invoked particle acceleration mechanism in Astrophysics is diffusive shock acceleration, also known as Fermi I process. Fermi\cite{fermia,fermib} initially envisioned a mechanism, later named Fermi II acceleration process, able to explain energy transfer from magnetized clouds to individual particles. The original idea was as follows: energetic particles are scattered by magnetic irregularities; in an interaction with a magnetized cloud the particle changes its direction of motion and gains or loses energy depending on whether the collision is head-on or tail-on; after averaging over many particle-cloud interactions, thanks to the fact that the head-on collisions turn out to be slightly more numerous, the process leads to a net energy gain $\sim (\Delta E/E) =(4/3)(V/c)^2$, where $V$ is the velocity of the clouds and $c$ is the speed of light. Since the velocity of the cloud, or of the magnetic perturbation, is generally $V\ll c$, in this formulation the acceleration turns out to be very slow. Things change dramatically, however, when the same process is considered in the context of a shock: the velocity of magnetic irregularities, which serve as scattering centres for particles, is negligible with respect to the fluid velocity on both sides of the shock. This implies that the scattering centres can be considered at rest with the fluid both upstream and downstream. In the shock system, whether a particle moves from upstream to downstream or viceversa, it always sees the fluid on the other side of the shock as approaching. Therefore, each time an energetic particle crosses the shock, it always suffers head on collisions with scatterers on the other side of the shock, thus gaining energy much faster: $(\Delta E/E)\propto (V_S/c)$, where $V_S$ is now the shock velocity (which is also much higher than the typical velocity of magnetic perturbations in the Galaxy). The fact that the energy gain is now linear in the velocity ratio is what gave origin to the designation ``Fermi I'', as opposed to ``Fermi II'', used to indicate the stochastic acceleration process in which the gain is second order in $V/c$. 

A very attractive feature of Fermi I mechanism is the fact that in the context of a strong shock it gives rise to a particle spectrum which is a power law in momentum with a universal slope, close to what implied from CR observations. In general the spectrum of shock accelerated particles will be given by $N(p)\propto p^{-\gamma_p}$, where $N(p)$ is the number of particles per unit momentum interval and 
\be
\gamma_p=3 R_T/(R_T-1)\ ,
\label{eq:slope}
\ee
with $R_T$ the compression ratio of the shock, namely $R_T=u_1/u_2$ with $u_1$ and $u_2$ the fluid velocities upstream and downstream of the shock respectively. 
The compression factor $r$ only depends on the shock Mach number $M_s=u_1/c_{s1}$ where $c_{s1} \approx 10 \sqrt{T_4}{\rm km/s}$ is the sound speed in the Interstellar Medium (ISM), whose temperature has been expressed in units of $10^4$ K. Using the standard Rankine-Hugoniot relations to describe the jump of all thermodynamical quantities at the shock, one finds:
\be
R_T=\frac{4 M_s^2}{3+M_s^2}\ .
\label{eq:shockr}
\ee 
For strong shocks, namely $M_s\gg 1$, this ratio is $R_T \approx 4$ and hence $\gamma_p=4$. For relativistic particles ($E\gg mc^2$), this slope in momentum is equivalent to a slope in energy $\gamma_e$ that is easily calculated using the fact that $E^{-\gamma_e}dE=4 \pi p^2 p^{-\gamma_p} dp$. The result is $\gamma_e=2$, which is exactly what required to explain the CR spectrum at energies below the {\it knee}, if propagation effects (to be discussed later) lead to a steepening by $\sim 0.7$. The spectral index will be $\gamma_e>2$ for weaker shocks.

A noticeable feature of this process is that the particle spectrum is completely insensitive to the scattering properties. This is because the probability for particles to return to the shock is unaffected by scattering. What does depend on scattering, however, is the time it takes for the particles to get back to the shock, and hence the maximum number of crossings a particle can undergo during the life-time of the system, or before being affected by energy losses: in other words, the maximum achievable energy. 
In the diffusive regime, the time it takes for a particle to complete a cycle around the shock is\cite{druryrev}:
\be
t_{\rm acc}=\frac{3}{u_1-u_2}\left[\frac{D_1}{u_1}+\frac{D_2}{u_2}\right]
\label{eq:tacc}
\ee
where $D_1$ and $D_2$ are the diffusion coefficients upstream and downstream of the shock, which depend on the particle energy and on the level of magnetic turbulence.
The maximum achievable energy is then determined by the condition that the acceleration time be less than the age of the system and the timescale for losses:
\be
t_{\rm acc}(E_{\rm Max})=\min\left(t_{\rm age}, t_{\rm loss}\right)\ .
\label{eq:maxen}
\ee
While losses are usually not a concern for protons, the system lifetime turns out to impose important constraints: SNR shocks remain efficient accelerators only for a relatively short time. Immediately after the SN explosion, the SN ejecta expand in the ISM with a velocity which is almost constant and highly supersonic. During this phase, the so-called free-expansion (but more properly ejecta-dominated) phase, acceleration is expected to be effective. After some time, however, the mass of ISM that the shock sweeps up becomes comparable to the mass of the ejecta, and, from that point on, the shock velocity starts to decrease. This happens at a time $T_{\rm ST}=R_{\rm ST}/V_{\rm ej}$, where $(1/2) M_{\rm ej} V_{\rm ej}^2=E_{\rm SN}$ and
$R_{\rm ST}$ is defined by the condition $(4/3) \pi \rho_{\rm ISM} R_{\rm ST}^3=M_{\rm ej}$. One finds:
\be
T_{\rm ST}=200 M_{\rm ej \odot}^{5/6} E_{51}^{-1/2} n_1^{-1/3} {\rm yr}\ ,
\label{eq:sttime}
\ee
where $E_{51}$ is the kinetic energy associated with the SN explosion in units of $10^{51}\ {\rm erg}$, $n_1$ is the ISM density in units of ${\rm cm}^{-3}$ and $M_{\rm ej \odot}$ is the mass of the ejecta expressed in units of solar masses. For typical values of the parameters the Sedov phase starts after few hundred years.

How this time compares with the acceleration time of energetic particles obviously depends on their diffusion coefficient (Eqs.~\ref{eq:tacc} and \ref{eq:maxen}). The latter can be worked out in a simple analytic way in the quasi-linear approximation, namely under the assumption that the perturbation $\delta B$ responsible for scattering is weak: $\delta B \ll B_0$, where $B_0$ is the regular field. If one studies the motion of a particle in this situation, one easily obtains that the particle scattering frequency is:
\be
\nu_s=\frac{\pi}{4} \Omega \left\langle \left( \frac{\delta B_{\rm res}}{B_0}\right)^2\right\rangle
\label{eq:scattnu}
\ee
where $\delta B_{\rm res}$ refers to the turbulent field at the resonant wavelength, namely at a wavelength equal to the particle Larmor radius $r_L$. Before writing the particle diffusion coefficient parallel to the regular field $B_0$, as $D=v \lambda_{\rm mfp}/3$, with $v$ the particle velocity and the mean free path $\lambda_{\rm mfp}=v/\nu_s$, let us introduce ${\cal F}(k)$, the power per logarithmic bandwidth at wavenumber $k$, such that:
\be
\left(\frac{\delta B(k)}{B_0}\right)^2=\int_0^{k} \frac{dk'}{k'}\ {\cal F}(k)\, \, \, \, \, {\rm and}\, \, \, \, \, \, \, \left(\frac{\delta B_{\rm res}}{B_0}\right)^2={\cal F}(1/r_L)\ .
\label{eq:dbf}
\ee
We then obtain for the diffusion coefficient\cite{blandeich}, in the case of a relativistic particle ($v=c$):
\be
D_\parallel(p)= \frac{c^2}{3\nu_s}=\frac{4}{3 \pi} \left(\frac{B_0}{\delta B_{\rm res}}\right)^2 c r_L\; \, \, \, \, \, 
D_\perp(p)=\left(\frac{r_L}{\lambda_{\rm mfp}}\right)^2 D_\parallel=\frac{\pi}{12} \left(\frac{\delta B_{\rm res}}{B_0}\right)^2 c r_L
\label{eq:qldiff}
\ee
where $_\parallel$ and $_\perp$ are with respect to the direction of the unperturbed magnetic field. One immediately sees that perpendicular diffusion is faster the larger the level of turbulence, while the opposite applies to parallel diffusion. These formulae are only strictly applicable, however, for $\delta B/B_0<1$, and for $\delta B \rightarrow B_0$ one has the so-called Bohm diffusion limit, with the particle mean free path becoming as small as their gyro radius:
\be
 D_\parallel \approx D_\perp \approx D_B=\frac{1}{3}cr_L\ .
\label{eq:dbohm}
\ee
Actually, determining the relative importance of parallel and perpendicular transport is a much more complicated task than suggested by Eq.~\ref{eq:qldiff}. This is due to the fact that the latter is actually dominated by magnetic field line wandering, which combines in a non-trivial way with parallel diffusion\cite{matbieber}. It is still true, however, that for small perturbations, one expects $D_\perp < D_\parallel$, and hence the most stringent constraints on the maximum achievable energy will come from parallel diffusion. 

Let us first consider the case in which $\delta B$ is just what can be expected assuming that turbulence in the Galaxy is injected with $\delta B/B_0\approx 1$ at a scale of $L_{\rm out}\sim 50-100$ pc and then develops a Kolmogorov spectrum at smaller scales. If $\delta B \sim B_0$ at the outer scale, then $\delta B_{\rm res}^2=B_0^2 \left(r_L/L_{\rm out}\right)^{2/3}$. As a result, the maximum achievable energy at the Sedov time is $E_{\rm Max}\sim$ a few GeV.

If instead, due to some other process, the field is amplified up to a level $\delta B_{\rm res}\approx B_0$, then $E_{\rm Max}\sim 10^4-10^5$ GeV\cite{lc83b}, which is still a factor 30-100 short of the {\it knee}.

The conclusion one reaches is then that the magnetic field in the vicinity of Supernova shocks must be amplified by a large factor compared to the average value in the ISM if SNRs are responsible of the acceleration of galactic CRs up to the {\it knee}. On the other hand, if SNRs are the main factories of Galactic CRs, another condition must also be true: of order 10\% of the kinetic energy of the blast wave must be converted into accelerated particles.This implies that the test-particle description of the shock acceleration process is not appropriate and the dynamical back reaction of the particles on the accelerators must be taken into account.

\section{Non-linear Diffusive Shock Acceleration}
\label{sec:nldsa}
The bottom line of the discussion in the previous section is that, if SNRs are the primary sources of Galactic CRs, a correct description of the acceleration process can only be achieved within the framework of Non-Linear Diffusive Shock Acceleration (NLDSA hereafter) with Magnetic Field Amplification (MFA hereafter). The full theory has been built in the course of several years with the contribution of several people and research groups using very different approaches. Pioneering work on the subject of CR modified shocks dates back to the early '80s\cite{druvolk81} and was based on two-fluid models: the thermal plasma and the CRs were treated as two distinct interacting fluids; no magnetic field amplification was included and magnetic fields played no dynamical role. Later on, in the early 90's, numerical approaches to the problem started to be developed, both finite difference schemes\cite{bereye94,berevolk97,berevolk00} and Montecarlo simulations\cite{ellisoneichler84,knerr96,kangjones97}. The numerical codes developed for this purpose have evolved during the years and have been constantly upgraded to take into account an increasing amount of physical processes, including now MFA\cite{bere04,volk05,vladielli08,vladielli09}.

\begin{figure}
\includegraphics[scale=.45]{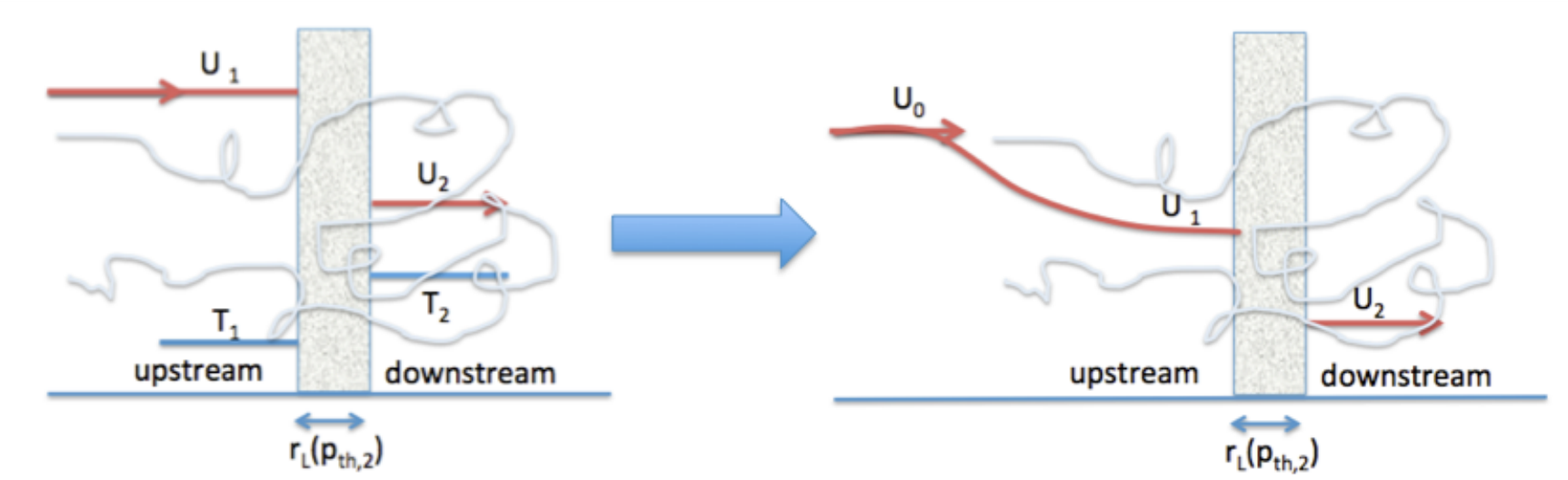}
\caption{Comparison between a shock that is not accelerating particles efficiently and one that is, and develops a precursor. In the left panel the standard picture of a test-particle shock is drawn: the fluid undergoes an abrupt transition, slowing down and being compressed over a scale comparable with the Larmor radius of thermal particles. In the right panel a sketch of a modified shock is shown: the fluid starts being slowed down and compressed over a scale comparable with the diffusion path length of the highest energy particles. Further compression by a factor that is always smaller than 4 occurs then on a scale comparable with the thermal particle Larmor radius, and this is the so called subshock (grey shaded area).}
\label{fig:shockmod}
\end{figure}

Also very fruitful have been the attempts at building a semi-analytical kinetic description of the system dynamics\cite{blasi02,ab05,ab06,bac07}. This kind of modelling has much evolved during the years as well, and, in its present form\cite{apjl08,esc09,bfeed09} it includes all three major sources of non-linearity:
\begin{itemize}
\item[1)]{the dynamical effect of the CR pressure on the shock}
\item[2)]{the wave generation by accelerated particles streaming upstream of the shock}
\item[3)]{the dynamical reaction of the amplified magnetic field on the system.}
\end{itemize}
Before presenting the equations, it is probably useful to give a qualitative description of what happens. In case of efficient acceleration, accelerated particles take on a sizeable fraction of the energy that is being dissipated at the shock. Energetic particles have velocities much larger than the shock speed and large mean free paths: they can propagate far from the shock in the upstream and carry with them information about the presence of the shock. In terms of equations, this fact is expressed by a CR related pressure term that is computed based on the transport equation for accelerated particles. This CR pressure term ($P_c$ in the following) is largest at the shock location and progressively decreases far upstream from the shock, due to the fact that progressively fewer particles can reach large distances. This is in turn a consequence of the diffusive transport of particles: they diffuse further the larger their energy (in all ``normal'' situations the diffusion coefficient is an increasing function of energy), and only the most energetic ones can reach very far upstream, with $P_c$ tending to zero at upstream infinity. As a result of the conservation of total momentum (and mass), a {\it precursor} then develops in the thermal plasma, namely a region in which the fluid progressively slows down while approaching the shock from upstream infinity, and at the same time its density increases. The shock develops a profile such as that represented in the right panel of Fig.~\ref{fig:shockmod}. The velocity jump between upstream infinity and downstream is realised in two steps, with an initial gradual decrease and then a real discontinuity ({\it i.e.} sudden variation of the thermodynamical quantities on a scale comparable with the Larmor radius of thermal downstream particles) at the so called subshock (shaded area in the right panel of Fig.~\ref{fig:shockmod}). While the compression at the subshock ($u_1/u_2$) is now less than 4 even for a large Mach number shock, the overall compression ratio, $u_0/u_2$, is now potentially very large, much larger than 4: this is due to the fact that CRs can escape the system (from upstream) and take away energy with them, so that the shock turns out to be effectively radiative.

An important consequence related to the development of a precursor is that the spectrum of accelerated particles is now no longer a power-law, but it rather develops a concave shape, with a slope in momentum, $\gamma_p$, that is steeper than 4 at low energies and flatter than 4 at high energies. This is easily understood based on the expression for $\gamma_p$ in Eq.~\ref{eq:slope} and the velocity profile in the right panel of Fig.~\ref{fig:shockmod}, that clearly shows how the velocity compression ratio is now a function of distance from the shock. Particles of different energies diffusively propagate to different distances from the shock, with lower energy particles confined closer to the shock than higher energy ones. As a consequence, the former experience only the compression ratio of the subshock, which is less than 4, making the spectrum steeper than $p^{-4}$, while the latter propagate into the precursor experiencing progressively larger compression ratios. Far enough in the precursor the compression ratio becomes $R_T>4$ and the spectrum flatter than $p^{-4}$.

\begin{figure}
\includegraphics[scale=.35]{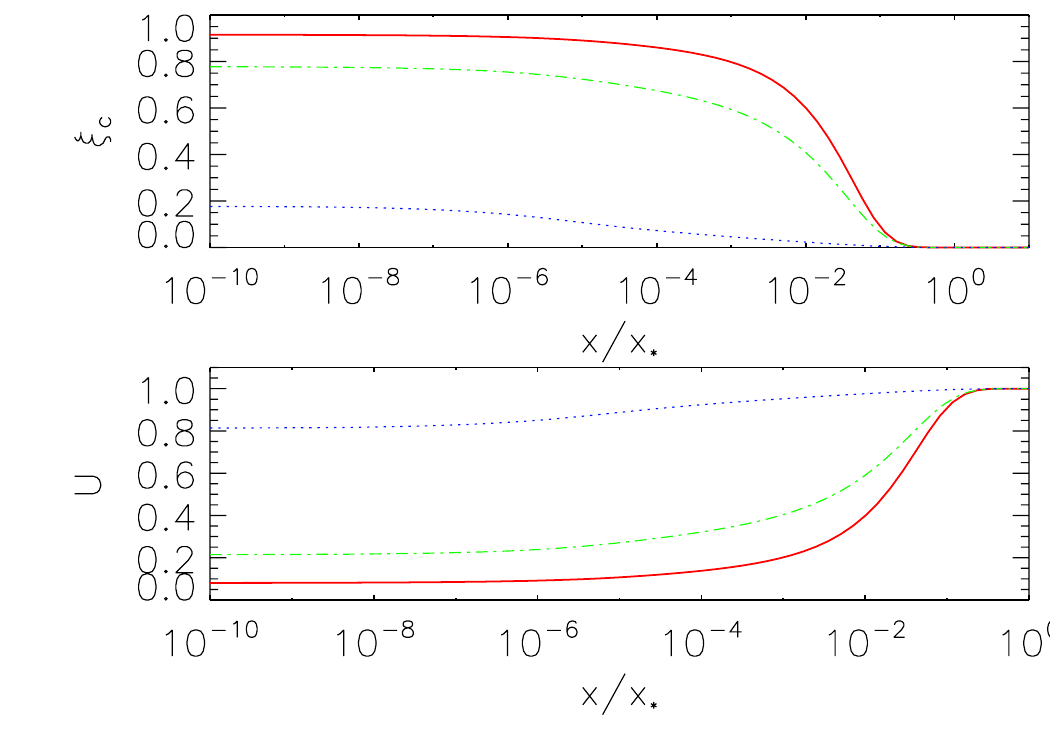}
\includegraphics[scale=.35]{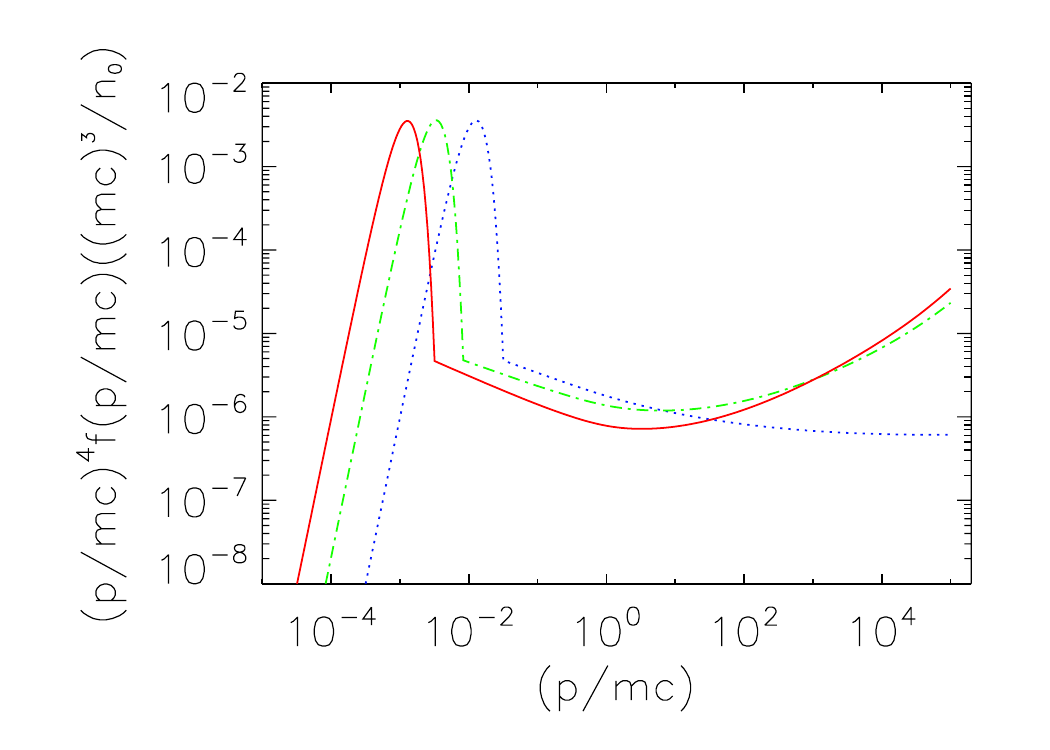}
\caption{Properties of a modified shock. The left panel shows the spatial profile of the fluid velocity and CR pressure in the region upstream of the shock. The shock position is coincident with the left boundary of the box. The upper panel shows the CR pressure normalised to the shock ram pressure. The lower panel shows the fluid velocity profile normalised to the velocity at upstream infinity. On the right we plot the particle spectrum immediately downstream of the shock. In all panels the different curves correspond to different Mach numbers: red (solid) is for $M_0=100$; green (dot-dashed) is for $M_0=30$, blue (dotted) is for $M_0=5$.}
\label{fig:allspec}
\end{figure}

The picture just described qualitatively, is illustrated in a more quantitative way in Fig.~\ref{fig:allspec}, where we show the cosmic ray pressure profile and fluid velocity profile on the left, and the particle spectrum immediately downstream on the right, all as a function of shock modification. Before discussing the results shown in this figure, let us consider the general set of equations to be solved in order to fully describe the modified shock dynamics: 
\begin{itemize}
\item{mass conservation}
\be
\frac{\partial \rho}{\partial t}=-\frac{\partial (\rho u)}{\partial x}
\label{eq:masscons}
\ee

\item{momentum conservation}
\be
\frac{\partial \rho}{\partial t}=
-\frac{\partial}{\partial x} \left[\rho u^2 + P_g + P_c + P_w\right]
\label{eq:momcons}
\ee

\item{energy conservation}
\be
\frac{\partial}{\partial t} \left[\frac{\rho u^2}{2}+\frac{P_g}{\gamma_g-1}\right]=-\frac{\partial}{\partial x} \left[ \frac{\rho u^3}{2}+\frac{\gamma_g P_g u}{\gamma_g-1}\right]-u\frac{\partial}{\partial x} \left[P_c+P_w\right]+\Gamma_{\rm th} E_w
\label{eq:encons}
\ee

\item{transport equation}
\begin{eqnarray}
\frac{\partial f(t,x,p)}{\partial t}+\hat u(x) \frac{\partial f(t,x,p)}{\partial x} &=& \\
&=&\frac{\partial}{\partial x}\left[D(x,p)\frac{\partial f(t,x,p)}{\partial x}\right]+\frac{p}{3} \frac{\partial f(t,x,p)}{\partial p} \frac{d \hat u(x)}{dx} \nonumber 
\label{eq:transp}
\end{eqnarray}

\item{evolution of the waves}
\be
\frac{\partial E_w}{\partial t}+\frac{\partial F_w}{\partial x} = u \frac{\partial P_w}{\partial x} + \sigma E_w-\Gamma_{\rm th} E_w
\label{eq:waveev}
\ee

\item{turbulent plasma heating}
\be
\frac{\partial P_g}{\partial t}+u \frac{\partial P_g}{\partial x} + \gamma_g P_g \frac{du}{dx}=(\gamma_g-1)\Gamma_{\rm th} E_w
\label{eq:heating}
\ee

\end{itemize}
In the above equations, the different symbols have the following meaning. $\rho$ is the gas density, $u$ its velocity, $P_g$ and $\gamma_g$ its pressure and adiabatic index respectively. $P_c$ is the CR pressure, to be derived from their distribution function $f$. In the transport equation $D$ is the diffusion coefficient and $\hat u$ is the velocity of the scattering centres: $\hat u(x)= u(x)\pm v_w(x)$, with $v_w$ the phase velocity of the waves. Finally $P_w$, $E_w$ and $F_w$ are the pressure, energy and energy flux of the magnetic waves, while $\sigma$ and $\Gamma_{\rm th}$ are the wave growth and damping rate respectively. The latter obviously coincides with the rate at which the plasma is heated non-adiabatically (Eq.~\ref{eq:heating}). 

The final missing ingredient is a recipe for injection of particles into the acceleration process. For that, a widely adopted prescription is the so-called thermal leakage\cite{bgv05}, namely the assumption that all particles with a Larmor radius larger than some multiple of the shock thickness start being accelerated. The latter is comparable in turn with the Larmor radius of the thermal particles downstream, so that in the end all particles with $p>p_{\rm inj}= \xi p_{\rm th,2}$ are injected, with $p_{th,2}=\sqrt{2 m_p k_B T_2}$. It is worth mentioning that while the thermal leakage has been a popular prescription for particle injection for a long time, recent results of hybrid simulations\cite{damtolik14} seem to point to picture in which the probability for a particle to start being accelerated is independent on its momentum. The detailed physics of injection is an important subject that can only be investigated by means of numerical simulations, with Particle In Cell and hybrid methods. These kinds of simulations, especially in more than 1 spatial dimension, were for a long time beyond the existing computational capabilities. While in the least decade, PIC simulations have been used intensively to investigate relativistic shocks (less challenging from the computational point of view), only now hybrid simulations start appearing in the literature\cite{damtolik14,damtolik13} with the set-up appropriate to describe non-relativistic shocks, bearing the promise of fundamental progress in our understanding of the acceleration physics at collisionless shocks.

A steady state solution for this set of equations in the shock frame can be found through an iterative method\cite{ab05,ab06,esc09,bfeed09}. In Fig.~\ref{fig:allspec}, we illustrate the result of such a calculation carried out for a shock velocity $u_0=5 \times 10^8 {\rm cm}{\rm s}^{-1}$ and assuming thermal leakage with $\xi=3.5$. We plot the predicted particle distribution function and the spatial profiles of fluid velocity and CR pressure. The results shown in this figure deserve some comments.

On the left panel, the precursor qualitatively illustrated in the right panel of Fig.~\ref{fig:shockmod}, now clearly appears to become more pronounced with increasing shock Mach number. This is due to the fact that the shock acceleration efficiency is increasing with $M_0$, as shown on the upper left panel of the figure, where the fraction of energy that is transformed in CRs, $\xi_c=P_c/(\rho v_S^2)$, is shown. With increasing $\xi_c$,  the CR spectrum becomes progressively more modified: its concavity becomes more pronounced and the high energy slope well flatter than 4. At the same time the peak of the maxwellian part of the distribution function progressively shifts towards the left: the larger fraction of energy converted into accelerated particles has been taken from the thermal energy of the downstream plasma, whose temperature decreases.
Overall, Fig.~\ref{fig:allspec} illustrates well two of the three most important predictions of the non-linear theory of shock acceleration: in a SNR that is efficiently accelerating CRs, one expects to measure a lower downstream plasma temperature than Rankine-Hugoniot relations would dictate, and, if direct signatures of non-thermal particles can be seen (e.g. through $\gamma$-ray emission associated with the decay of pions from nuclear scatterings), their high energy spectrum should be flatter than $p^{-4}$.

In addition to these two, there is a third signature of efficient acceleration, namely the amplification of magnetic fields induced by the accelerated particles. We already mentioned that accelerated particles can stream ahead of the shock, which is exactly how a precursor is created. What we should add, however, is that this streaming provides a source of instability in the upstream plasma that leads to magnetic field amplification. This process might produce magnetic fields which are much more intense than the average $B_0$ in the ISM and this is the reason why we are forced to include magnetic pressure and energy terms in the conservation equations (Eqs.~\ref{eq:momcons}-\ref{eq:encons}): the magnetic pressure in the upstream can easily grow to values that exceed the thermal pressure of the plasma\cite{apjl08,bfeed09} and have important dynamical consequences. Indeed, when this happens, the plasma compressibility changes and as a result the shock modification is much reduced even for high Mach number shocks and high acceleration efficiencies, thus changing somewhat the results illustrated in Fig.~\ref{fig:allspec}, that was obtained ignoring these effects. In addition, as shown in Eq.~\ref{eq:heating}, part of the magnetic energy will be damped on to the plasma at the end of the turbulent cascade, increasing its temperature. This is the so-called turbulent heating\cite{mckenzie}. An accurate physical description of this phenomenon can be rather complex and uncertain, involving the microphysics of wave dissipation in a plasma, so usually it is simply accounted for through a phenomenological parametrization. The effects of turbulent heating will be especially important when discussing the properties of shocks propagating in a partially ionised plasma (Sec.~\ref{sec:neutrals}).

The physical process of magnetic field amplification (MFA hereafter) has received very much attention during the last few years, especially because, as we already mentioned, it provides our best hope to explain CR acceleration up to the {\it knee} in SNRs, possibly the only hope unless most SNR shocks are quasi-perpendicular. This very important topic will be introduced in the next subsection and then further discussed in Sec.~\ref{sec:mfanew}.

\subsection{Magnetic field amplification in the standard picture}
\label{sec:mfastd}
The fact that super-alfv\'enic streaming of CRs induces the growth of magnetic waves (Alfv\'en waves) has been well known for many years. Indeed, already in the '70s\cite{wentzel74} the interaction of CRs with self-generated Alfv\'en waves was suggested as the explanation for the observed near-isotropy of the CR flux. The general idea is as follows: as soon as there are energetic particles streaming at a speed that is larger than the local Alfv\'en speed (the characteristic speed of magnetic disturbances parallel to the pre-existing magnetic field in a medium), Alfv\'en waves of appropriate wavelength become unstable and grow in amplitude. The ``appropriate wavelength'' $2 \pi/\kappa_r$, with $\kappa_r$ the wavenumber, is established by the condition of resonance with the streaming particles' Larmor radius: $\kappa_r r_L=1$. The particles transfer momentum to the waves and the process leads to isotropization of the particle distribution. For relativistic particles, the residual anisotropy, after a few growth periods of the waves, is of order $v_A/c$, in excellent agreement with what direct observations of galactic CRs show\cite{amenomori05}. 

The waves excited when CRs propagate in the Galaxy have a well known growth rate, given by\cite{zweibel79,blandeich}: 
\be
\Gamma_{\rm res}=\frac{\pi}{8} \frac{v_d}{v_A}\frac{n_{\rm CR}(p>p_{\rm res})}{n_i} \Omega_{ci}^*\ , 
\label{eq:stdrate}
\ee
where $v_A=\sqrt{B_0^2/(4 \pi n_i m_i)}$ is the Alfv\'en speed with $B_0$ the unperturbed magnetic field and $m_i$ and $n_i$ the mass and density of ions in the medium, respectively, $\Omega_{ci}^*=e B_0/(m_i c)$ is the ion cyclotron frequency and $v_d$ is the particle drift velocity. 

When one is concerned with magnetic field amplification associated with particle acceleration in a SNR, $v_d$ coincides with the shock velocity. From the expression for $\Gamma_{\rm res}$ in Eq.~\ref{eq:stdrate}, one can derive the wave energy growth rate $\sigma$ that appears in Eq.~\ref{eq:waveev}, as $\sigma=2 \Gamma_{\rm res}$. 

In addition, the relation between energy flux, energy and pressure are all well known for Alfv\'en waves and so are their transmission properties. In practice, in the case of Alfv\'en waves, Eqs.~\ref{eq:masscons}-\ref{eq:heating} can be rephrased in such a way as to form a closed set, in which also the amplified magnetic field, the particle diffusion coefficient\cite{ab06} and the maximum achievable energy\cite{bac07} can be calculated self-consistently as a function of the shock speed $u_0$, thermal leakage parameter $\xi$ and turbulent heating parameter $\Gamma_{\rm th}$. 

The calculation can be performed at different stages during the evolution of a SNR, assuming that a steady-state is always reached on a time-scale that is less than the dynamical time-scale of the remnant. Under this assumption, which is expected to be a good one except for the highest energy particles, both the spectrum of accelerated particles in the source and the flux of particles escaping from the source can be computed as a function of time. Let us briefly discuss what the resulting spectrum of CRs injected in the ISM turns out to be.

\subsection{Escaping particles}
\label{sec:escape}
The escape of particles towards upstream infinity during the acceleration process is an essential feature of models of non-linear CR acceleration at shocks: as we already mentioned, this phenomenon is the very reason why compression factors larger than 4 can be achieved in efficient accelerators. Of course, aside from the theory one adopts to describe the accelerator, effective escape while the acceleration is still ongoing is fundamental if high energy particles must be released in the ISM. Waiting for the shock to disappear and for the SNR to merge in the ISM would imply large adiabatic losses for the particles and CRs at the {\it knee}, that are so difficult to accelerate, would never be detected at Earth. 

Understanding how particles escape from the shock is obviously essential to understand how the spectrum of CRs that we observe at Earth is formed. The system of Equations in Sec.~\ref{sec:nldsa} automatically tells us what fraction of the energy flux is carried away by escaping particles. What it does not tell us immediately is their spectrum. This can however be worked out\cite{esc09,freeesc10,spec10} and the result is that, at each time during the system evolution, it consists of a function strongly peaked in energy around the maximum energy that the system is able to guarantee at that time. If this is the case, then one can give a back of the envelope estimate of the spectrum released over time, simply by expressing the energy flux carried away by the particles at $p_{\rm max}$ to a fraction $f_{\rm esc}$ of the total energy flux processed by the shock:
\be
4 \pi p^2 N_{\rm esc}(p)  c\ p\ dp=f_{\rm esc} \frac{1}{2} \rho v_S^3 4 \pi R_S^2 dt
\label{eq:escspec}
\ee
Assuming $R_S\propto t^\alpha$ (and hence $V_S \propto t^{\alpha-1}$) and assuming that $p_{\rm max}\propto t^\beta$ (as would be true for example if the maximum momentum were limited by the size of the system, $p_{\rm max}\propto R_S$) it is straightforward to see, that, since $dt/dp\propto t/p$, one obtains, from a SNR expanding in the uniform ISM:
\be
N_{\rm esc}(p)\propto f_{\rm esc}\ p^{-4}\ t^{5 \alpha-2} 
\label{eq:escfin}
\ee

\begin{figure}
 \begin{center}
 \includegraphics[scale=.5]{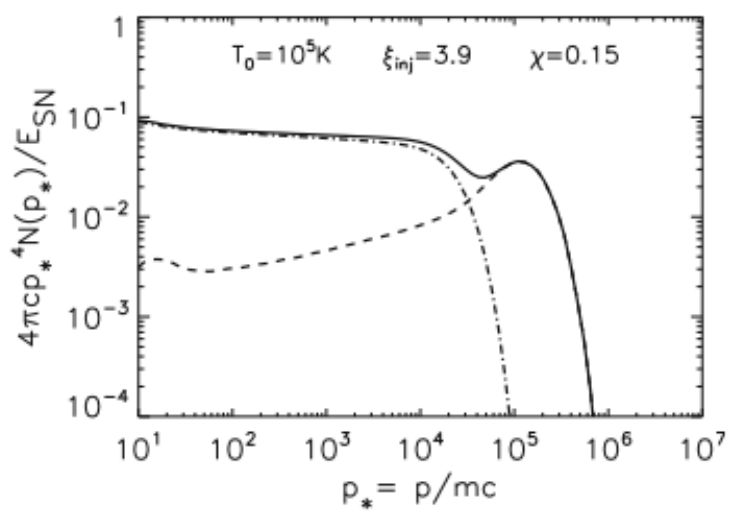}
 \end{center}
 \caption{Left panel: the spectrum of CRs released in the ISM (Ref.~\protect\refcite{spec10}\protect) by a SNR that accelerates particles efficiently during its lifetime, while expanding in a uniform medium with temperature $T=10^5$ K. The dashed line is the spectrum of particles escaping from upstream; the dot-dashed line is the spectrum of particles that are advocated in the downstream and released only at late times, when the SNR merges in the ISM.
 }
 \label{fig:escspec}
 \end{figure}

It follows, that in the Sedov-Taylor phase, when $\alpha=2/5$, SNRs will release a spectrum $N_{\rm esc}(p)\propto p^{-4}$, if $f_{\rm esc}$ keeps constant with time. Since, during the Sedov-Taylor phase, we expect $\beta<0$, namely the maximum momentum to decrease with time, in order to have a spectrum steeper than $p^{-4}$, the fraction of energy that escaping particles take away, $f_{\rm esc}$ would have to increase with time. In reality, when the entire calculation, including acceleration and escape, is carried out self-consistently, the spectrum of escaping particles often turns out to be flatter than $p^{-4}$. In Fig.~\ref{fig:escspec} we show the result of such a calculation , where the maximum energy at every time is set by the condition that particles can only be accelerated up to energies such that their Larmor radius does not exceed 15\% of the SNR radius at that time. Three important features can be noticed in the Figure, that corresponds to the particle release by an ``average'' SNR, that accelerates particles rather efficiently while evolving in a uniform medium with temperature $10^5$ K: 1) while the spectrum of escaping particles is flatter than $p^{-4}$, the overall spectrum (escaping particles plus particles advocated in the downstream and released at late times) has a slope close to $p^{-4}$; 2) only particles that escape the system from upstream (dashed line) contribute at the highest energies; 3) the maximum energy that can be achieved by such remnant falls short of the {\it knee}.

\section{Theory confronts observations}
\label{sec:thvsdata}
The discussion above was meant at illustrating NLDSA theory and its basic predictions. The first obvious test of the association between Galactic CRs and SNRs involves making predictions for signatures of efficient acceleration in these systems, and checking them against the data. The most obvious place to start from is non-thermal emission. Several SNRs show non-thermal X-ray emission due to synchrotron radiation by relativistic electrons, and a few also show $\gamma$-ray emission, that usually can be due either to Inverse Compton scattering of the same relativistic electrons responsible for the X-rays, or to $\pi^0$ decay following nuclear collisions of relativistic protons: as we will discuss in the following detailed calculations taking into account multi-wavelength data are usually necessary in order to distinguish between these two scenarios.

In general, at a SNR shock that is efficiently accelerating CRs we expect to observe:
\begin{itemize}
\item[1.]{compression ratios larger than 4}
\item[2.]{downstream temperature less than what expected based on the Rankine-Hugoniot relations}
\item[3.]{concave particle spectra}
\item[4.]{possible signatures of the existence of a precursor, in the spatial structure of non-thermal emission}
\item[5.]{signatures of amplified magnetic fields}
\item[6.]{$\gamma$-ray observations: direct signatures of the presence of relativistic protons, up to the {\it knee}, through $\pi^0$ decay $\gamma$-rays where the target density for inelastic nuclear collisions is large enough}
\end{itemize}

A few relevant observations follow for each of the items listed above.

\subsection{Compression ratios}
We have indeed evidence in at least 2 young SNRs, Tycho and SN1006\cite{warren05,gamil07}. The evidence comes from measurements of the distance between the contact discontinuity and the shock, that leads to infer a compression ratio of order 7 in both cases. This value of the compression ratio is in perfect agreement with the predictions of NLDSA for the case of a shock that is efficiently accelerating particles and in which, either efficient turbulent heating takes place in the precursor, or the magnetic field is amplified to levels that make its energy density comparable with that of the thermal plasma upstream\cite{apjl08,bfeed09}.

\subsection{Downstream temperature}
An unexpectedly low downstream temperature has indeed been inferred in SNR RCW86 from Balmer emission\cite{helder09}. This measurement initially led its authors to deduce extremely efficient CR acceleration in this remnant, with conversion into accelerated particles of $\approx$ 50\% of the shock kinetic energy. This estimate then had to be revised, as we will discuss in Sec.~\ref{sec:neutrals}, but nonetheless, current estimates still give an acceleration efficiency about 20-30 \%. This subject will be discussed further in Sec.~\ref{sec:neutrals}.

\subsection{Concave spectra}
Also hints of concave spectra have been found: a few SNRs show a radio emission spectrum that becomes harder with increasing frequency\cite{reyelli92}. In addition a concave spectrum seems also favoured from fitting the SED of SN1006 and RCW86\cite{vink12}.

\subsection{Spatial profile of the emission}
SN1006 is also a good example of how the spatial profile of the emission can be used to derive the CR acceleration efficiency: indeed the X-ray emission profile is best reproduced\cite{rcw86} by assuming that the shock is accelerating CRs with an efficiency of order 30\%. The same exercise, carried out for the case of Tycho\cite{giodam12}, and considering both the X-ray and the radio emission, leads to conclude that here the acceleration efficiency is of order 10\%.

\subsection{Amplified magnetic fields}
In the last few years {\it Chandra} has allowed us to measure the thickness of the X-ray emitting region, showing that in a number of remnants this is extremely compact, of order of $0.01\ pc$. The simplest interpretation of these thin rims of emission is in terms of synchrotron burn-off: the emission region is thin because electrons lose energy over a scale that is of order $\sqrt{D \tau_{\rm sync}}$, where $D$ is the diffusion coefficient and $\tau_{\rm sync}$ is their synchrotron lifetime. Assuming Bohm diffusion, this length turns out to be independent of the particle energies and given by $\sqrt{D \tau_{\rm sync}}\approx 0.04 B_{-4}^{-3/2}$, where $B_{-4}$ is the field in units of 100 $\mu G$, therefore requiring that the magnetic field responsible for both propagation and losses be in the 100 $\mu G$ range\cite{vink12,blasirev}. 

Another observation that led to infer a large magnetic field is that of fast time-variability of the X-ray emission in SNR RX J1713.7-3946\cite{uchirxj}. Again a field in the $100 \mu G-1 mG$ range was estimated, interpreting the variability time-scale as the time-scale for synchrotron losses of the emitting electrons. 

Such high fields are strongly suggestive of efficient acceleration and of the development of related instabilities. However it should be mentioned that also alternative interpretations are possible\cite{bykovrev,schurerev}. For example their origin might be associated to fluid instabilities that are totally unrelated to accelerated particles\cite{giacajo}. Therefore while the evidence for largely amplified fields seems very strong, it cannot be considered as a definite proof of efficient CR acceleration.

In reality, the situation is even more complicated than so far described: recent re-analyses of the streaming instability at CR efficiency shocks show that its growth must be revised with respect to the estimate given in Eq.~\ref{eq:stdrate} and widely used in the modelling of NLDSA with MFA. In particular, at a shock where CRs are important in terms of the current they carry, the growth rate of resonant Alfv\'en waves is slower than described by Eq.~\ref{eq:stdrate}, whereas previously ignored non-resonant modes grow much faster. We will discuss this fact and its consequences in a dedicated Section (Sec.~\ref{sec:mfanew}).

\subsection{$\gamma$-ray emission}
The diffusion coefficient in the amplified field is sufficiently reduced so as to guarantee particle acceleration up to the {\it knee} in the early phases of SNR evolution. We would then expect to find some remnants accelerating particles up to energies close to 1 PeV among the young galactic SNRs. The high sensitivity of the latest generation of $\gamma$-ray telescopes, both from space and from the ground, has finally allowed us to observe the $\gamma$-ray emission of a number of remnants, in a quest for direct signatures of the presence of accelerated protons. These would reveal themselves through the process of pion production and successive decay: nuclear collisions of energetic protons with ambient gas produce pions, both charged and neutrals; the latter then decay into gamma-rays. In the environment of a SNR, however, even when $\gamma$-ray emission is detected, it is not straightforward to pin down its origin beyond doubt. In general $\gamma$-rays can be either of hadronic or leptonic origin (Inverse Compton Scattering of relativistic electrons responsible of lower frequency synchrotron emission). Multifrequency modelling of the emission can help disentangle between different scenarios, but firm conclusions are difficult to obtain. 

A good example of how complex the situation can be is provided by the case of SNR RX J1713.7-3946, the first SNR to be observed in TeV $\gamma$-rays\cite{aharonianrxj}. X-ray observations of this young remnant had pointed to the presence of a magnetic field in the $100\mu G$ range, suggestive of efficient CR acceleration\cite{rxjgio}. Therefore, when it was finally observed in high energy $\gamma$-rays, many different groups tried to model its multifrequency emission in order to understand whether the gamma-rays were most likely of leptonic or hadronic origin\cite{rxjgio,berevolkrxj}: the general conclusion was that the hadronic origin was most likely. Then more refined modelling of thermal X-ray emission started to cast doubts on this interpretation pointing towards a density of the ambient medium lower than required by the hadronic models\cite{ellirxj}. In the end, the initial conclusion had to change radically when lower energy gamma-ray observations with {\it Fermi}\cite{fermirxj}, showed that the photon spectrum in the MeV-GeV energy range was too flat to be explained as due to $\pi^0$ decay.

\begin{figure}
\includegraphics[scale=.35]{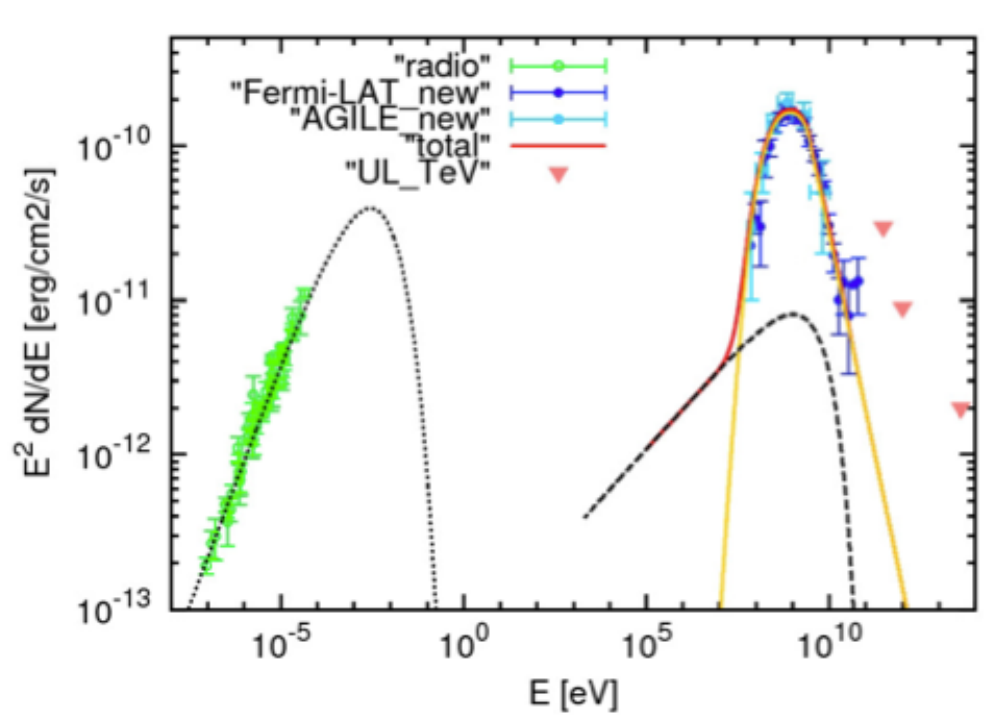}
\includegraphics[scale=.15]{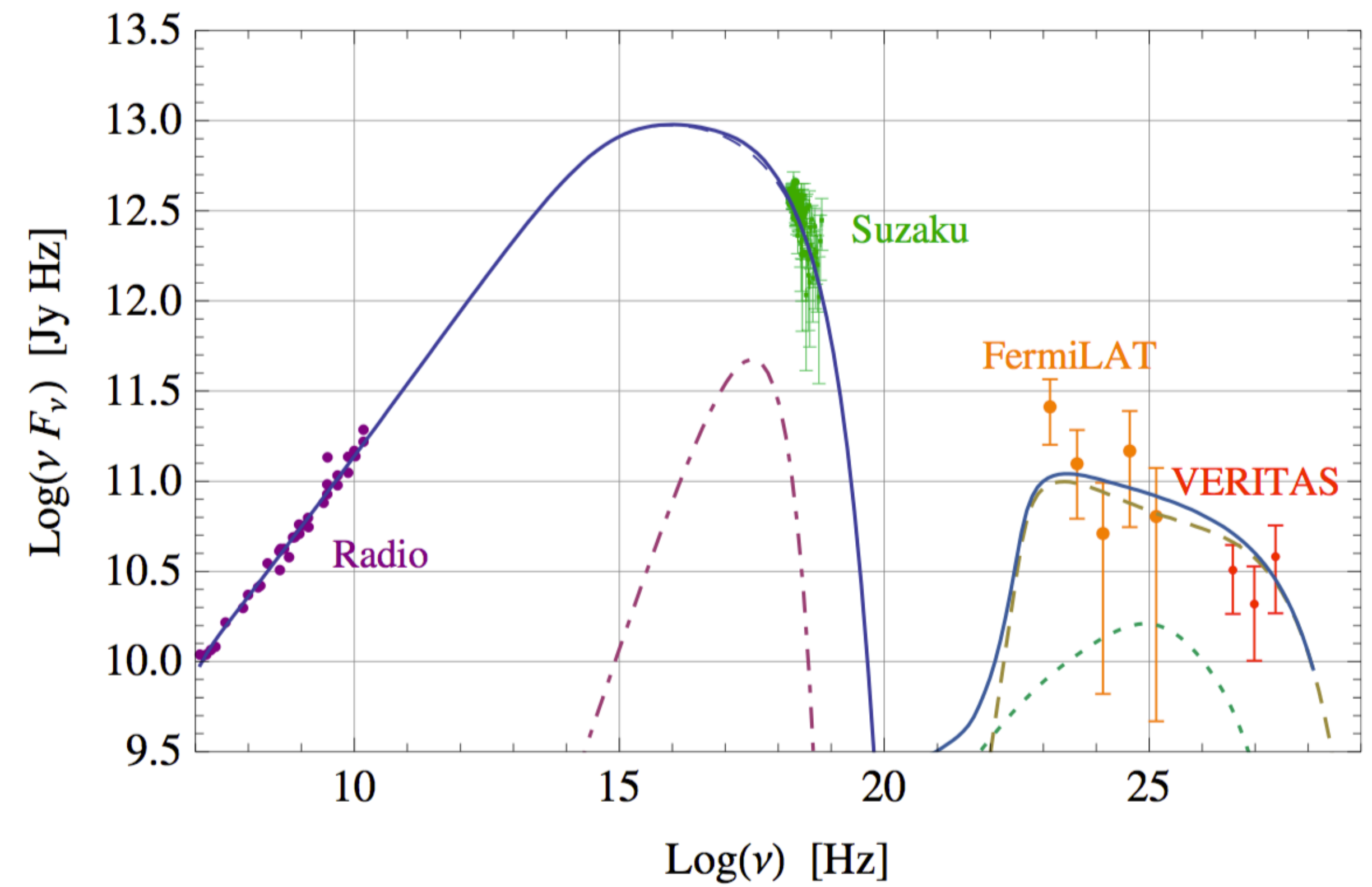}
\caption{Left panel: the integrated emission spectra of W44 (Ref.~\protect\refcite{w44martina}\protect). Right panel: the integrated emission spectrum of Tycho (Ref.~\protect\refcite{giodam12}\protect). The different curves represent different contributions to the total emission. A detailed explanation can be found in Refs.~\protect\refcite{w44martina}\protect and \protect\refcite{giodam12}\protect, from which the figures are taken. In both cases the $\gamma$-ray emission is primarily of hadronic origin: yellow curve in the left panel, free dot-dashed curve one the right panel. In the first case the proton spectrum becomes very steep above 20 GeV, whereas in the second case the maximum proton energy is 500 TeV.}
\label{fig:w44}
\end{figure}

In recent times direct evidence of $\pi^0$ decay gamma-rays has been found in at least two middle aged SNRs interacting with molecular clouds, W44\cite{w44fermi,w44agile} (whose broad band spectrum is shown in the left panel of Fig.~\ref{fig:w44}) and IC443\cite{ic443agile,ic443fermi}. For both remnants the hadronic interpretation of the gamma-ray emission is considered as certain, so we finally face the long-sought evidence of accelerated protons in SNRs. While this discovery provides support to the CR-SNR paradigm, it does not really show us the CR accelerators at their best. We mentioned that these are middle-aged SNRs (tens of thousands year old) and they are likely not accelerating CRs very efficiently: their $\gamma$-ray brightness is not due to a high flux of relativistic hadrons, but rather to a high target density for nuclear collisions, thanks to the interaction with a molecular cloud. And indeed, when one takes a closer look at the gamma-ray spectrum, the striking feature common to both remnants is an inferred proton spectrum that is steeper than $E^{-2}$ at all energies and becomes very steep ($E^{-\gamma_e}$ with $\gamma_e\approx2.5-3$) at energies larger than a few tens of GeV. Both features are very different than what we were expecting to see in the sources of Galactic CRs (flat spectra and $10^4$ times larger proton energies).

Direct proof of efficient CR acceleration in SNRs can only be found in young objects, in spite of the fact that their modelling might be more complicated and definite conclusions more difficult to draw, as we already mentioned for the case of RX J1713.7-3946. Currently, the best candidate as a hadronic emitter among young remnants seems to be Tycho (whose broad band spectrum is shown in the right panel of Fig.~\ref{fig:w44}). Here, multifrequency modelling\cite{giodam12} leads to infer a maximum proton energy of order 500 TeV, which is definitely getting close to the {\it knee}. However, even in this case the picture is not fully satisfactory: the radiation spectrum implies a spectrum of the emitting particles that is $f\propto E^{-\gamma_e}$ with $\gamma_e\approx2.2-2.3$. Again the observations point to a particle spectrum that is steeper than $E^{-2}$, contrary to our theoretical expectations. An analogously steep spectrum is inferred for another young remnant, Cas A\cite{zirakashvilicasa}, although here the modelling is more complicated and the evidence that $\gamma$-rays are of hadronic nature less compelling.

\subsection{A succesfull comparison?}
Theory seems to come out well from the above comparison with observations if one just counts the number of successful predictions versus discrepancies. However the discrepancy in terms of particle spectra is clearly very serious. 

In addition, it is made even more serious by recent studies of the CR propagation in the Galaxy, that also seem to require CRs to be injected in the Galaxy with a spectrum steeper than $E^{-2}$. We review this latter result in the following, and then discuss possible ways in which the theory can be modified to account for steep source spectra.

\section{The CR spectrum injected in the Galaxy}
\label{sec:prop} 
The CR spectrum at Earth results from the combination of injection and propagation. The basic expectation for how the spectrum at Earth relates to that injected by the sources is easily obtained in the so-called leaky box model of CR propagation. In these models the Galaxy is described as a cylinder of radius $R_d$ and height $H$, with $R_d\approx 15$ kpc, the radius of the Galactic disk, and $H\approx 3$ kpc, the extent above the disk of the magnetised Galactic halo as estimated from radio synchrotron emission. Cosmic Rays are confined within this cylinder for a time $\tau_{\rm esc}\approx H^2/D(E)$ with $D(E)$ the diffusion coefficient in the Galaxy. Let us write the latter as $D(E)=D_0 E^\delta$, if CR sources inject a spectrum $N_s(E)\propto E^{-\gamma_{\rm inj}}$ the spectrum of primary CRs at Earth will be:
\be
N(E)\approx \frac{N_s(E) {\cal R}_{\rm SN}} {2 \pi R_d^2 H \tau_{\rm esc}} \propto E^{-\gamma_{\rm inj}-\delta}\ .
\label{eq:CRspec}
\ee
Therefore what we measure at Earth only provides us with the sum of $\gamma_{\rm inj}$ and $\delta$.
On the other hand, during their propagation in the Galaxy CRs undergo spallation processes and there are chemical elements that mostly result from these interactions, such as for example Boron. The spectrum of secondaries will be given by:
\be
N_{\rm SEC}(E)\approx N(E) {\cal R}_{\rm spall} \tau_{\rm esc}\propto E^{-\gamma_{\rm inj}-2 \delta}\ ,
\label{eq:secspec}
\ee
where ${\cal R}_{\rm spall}$ is the rate of spallation reactions. It is clear then that the ratio between the flux of secondaries and primaries at a given energy $N_{\rm SEC}(E)/N(E)\propto E^{-\delta}$ can provide us with a direct probe on the energy dependence of the Galactic diffusion coefficient and hence allow us to infer the spectrum injected by the sources. 

\begin{figure}
\includegraphics[scale=.5]{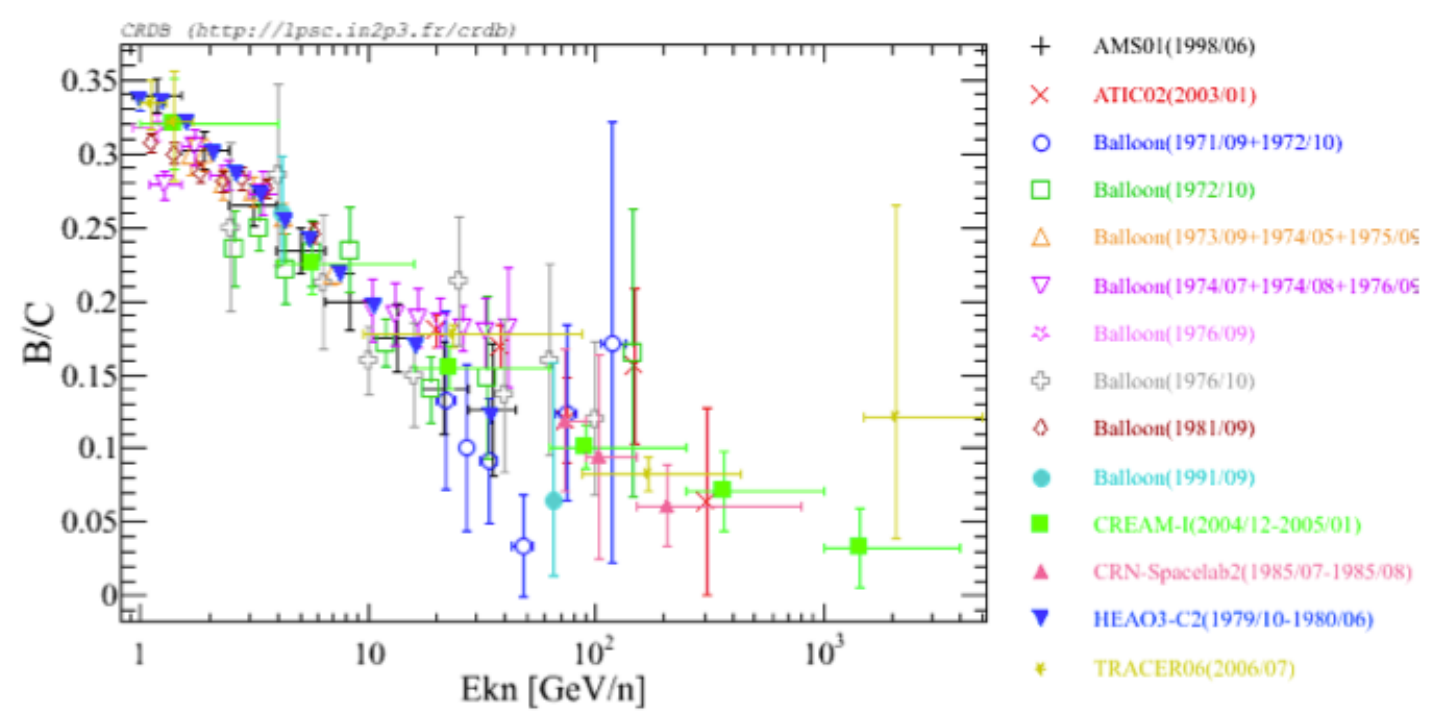}
\caption{B/C ratio as a function of energy per nucleon. Data from the Cosmic Ray Database (Ref.~\protect\refcite{maurinetal13}\protect).}
\label{fig:bcrat}
\end{figure}
 
A compilation of available measurements of the B/C ratio is shown in Fig.~\ref{fig:bcrat} as a function of energy per nucleon. It is immediately apparent from the Figure that the error bars on the high energy data points are rather large, and leave a considerable uncertainty on the energy dependence of the diffusion coefficient, being compatible\cite{deltaval} with anything in the interval $1/3<\delta<0.7$. As a consequence, the slope of the CR spectrum at injection is also uncertain in the interval $2<\gamma_{\rm inj}<2.4$.

\begin{figure}
\includegraphics[scale=.28]{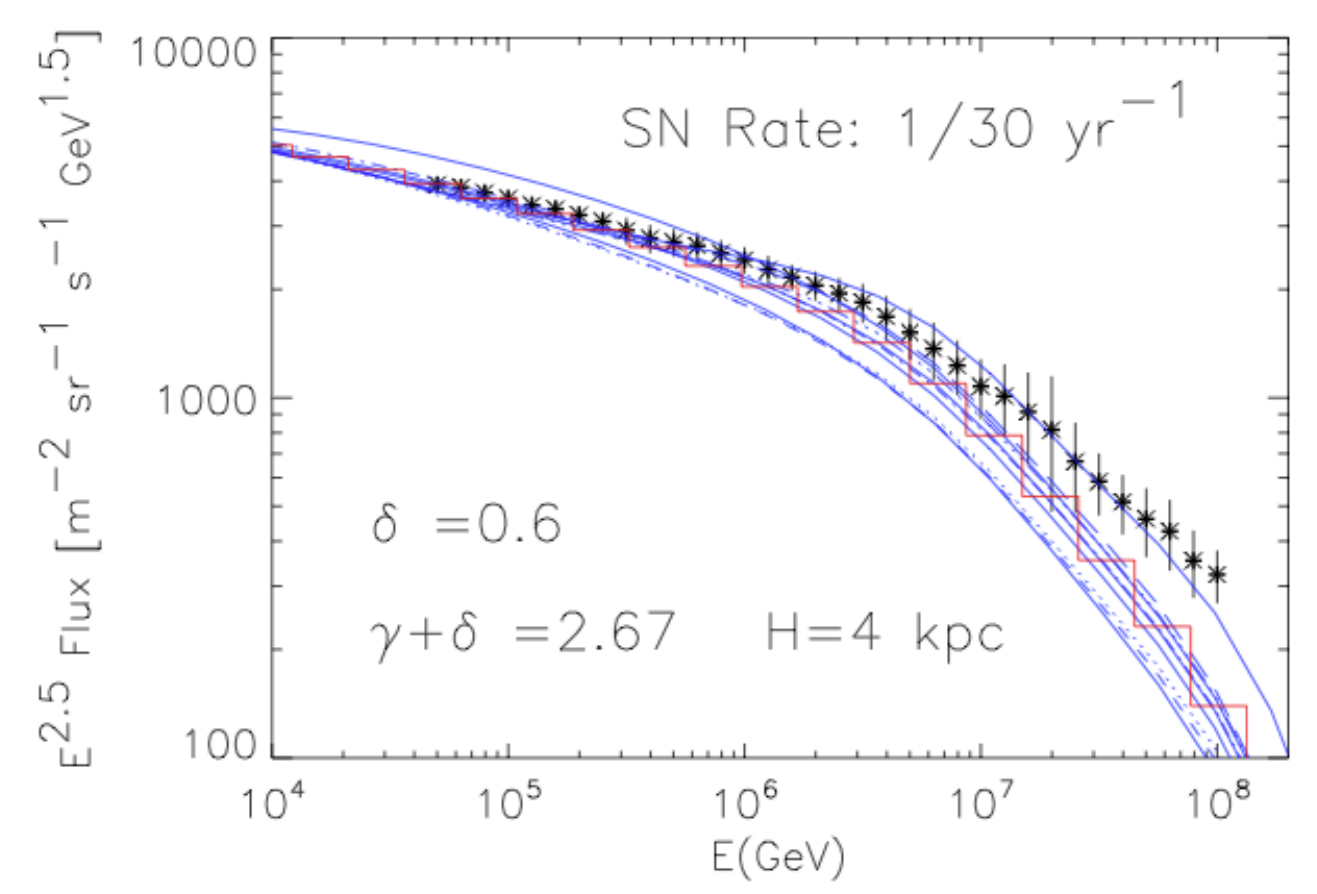}
\includegraphics[scale=.28]{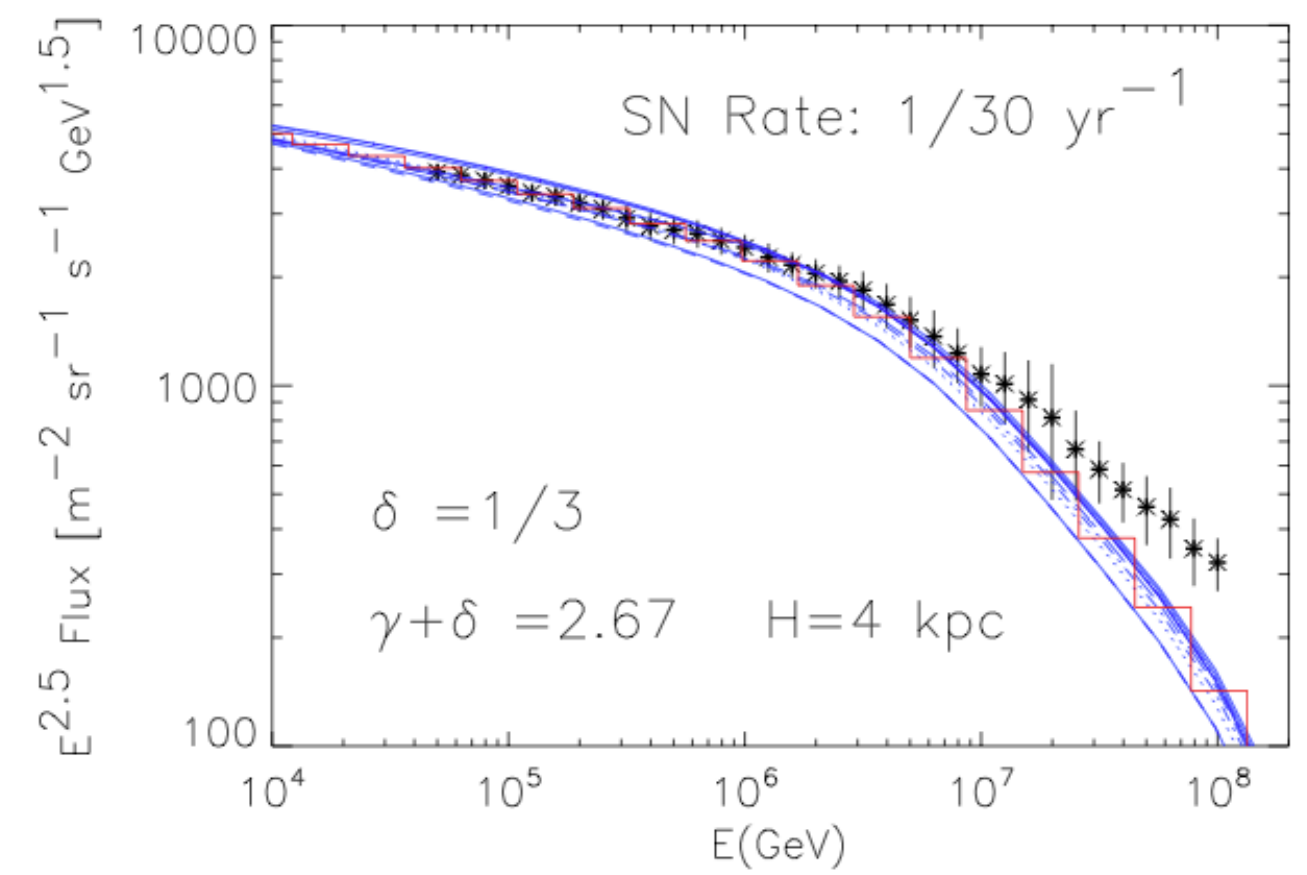}
\caption{All particle spectrum at Earth obtained from discrete sources under two different assumptions on injection and propagation (see text for more details). On the left $\gamma_{\rm inj}=2.34$ and $\delta=1/3$ is assumed, while on the right $\gamma_{\rm inj}=2.07$ and $\delta=0.6$. The symbols represent the data obtained as an average between the different experiments (Ref.~\protect\refcite{horandel}\protect); the different curves correspond to different realisations of the source distributions and the staircase like curve represents the average over the different realisations.}
\label{fig:allpart}
\end{figure}

\begin{figure}
\includegraphics[scale=.28]{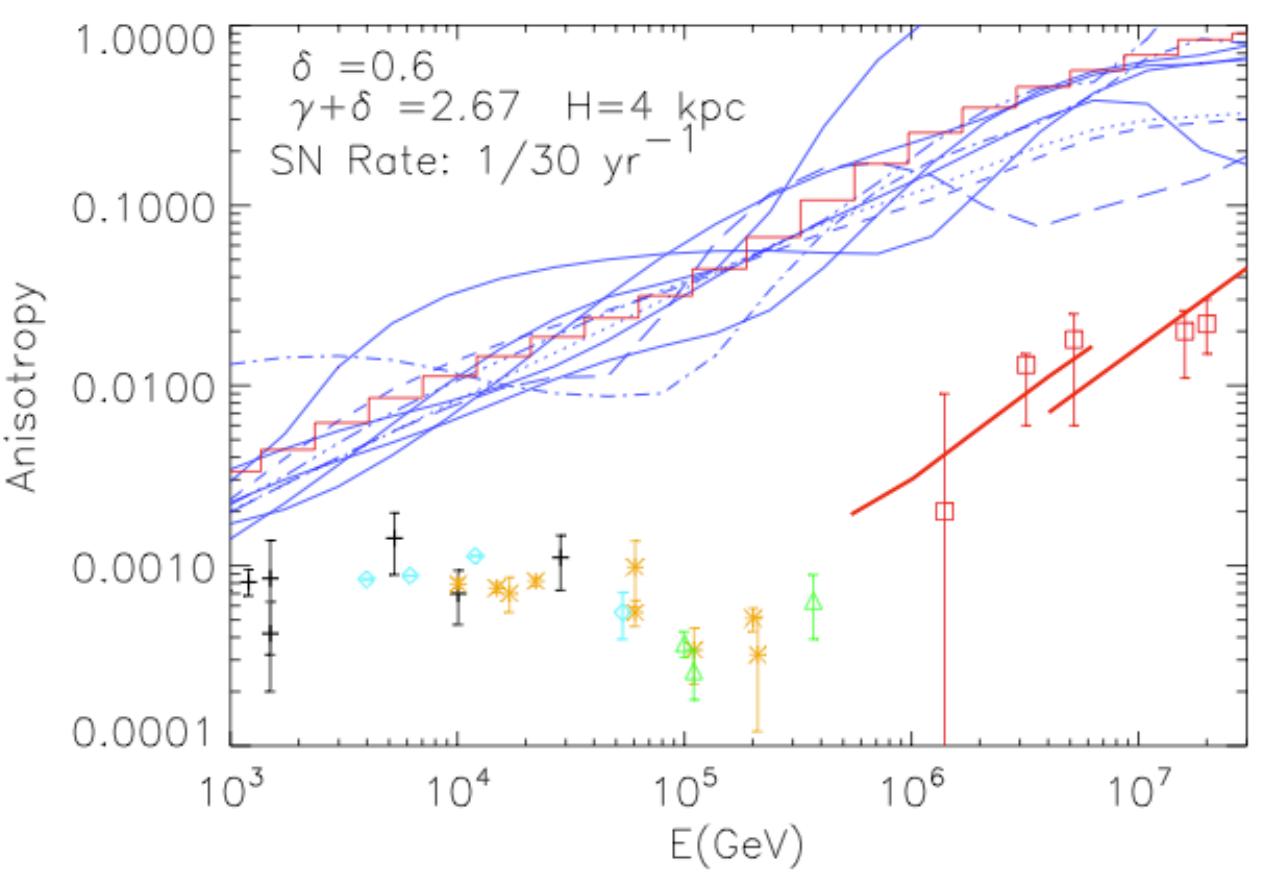}
\includegraphics[scale=.28]{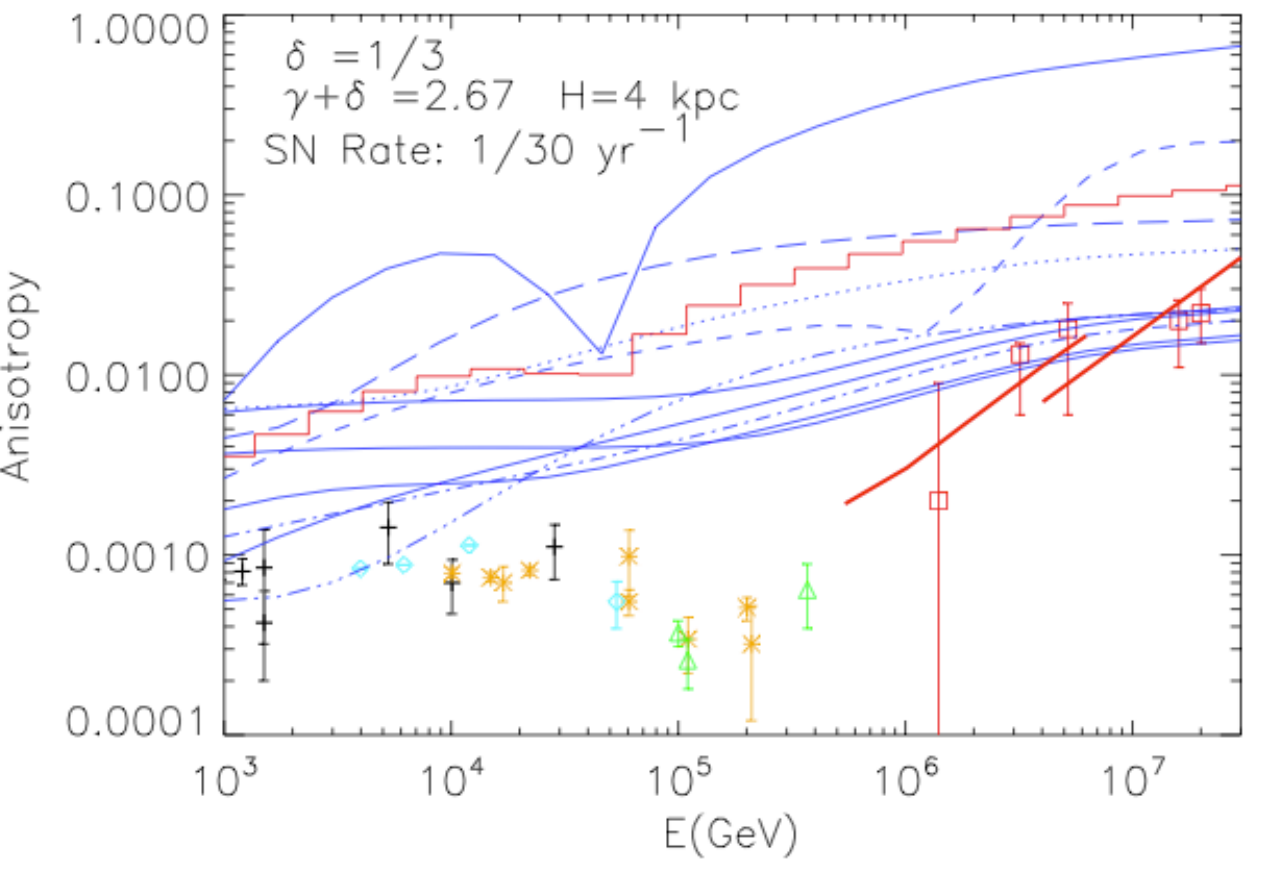}
\caption{Anisotropy of the CR flux at Earth obtained from discrete sources under two different assumptions on injection and propagation (see text for more details). On the left $\gamma_{\rm inj}=2.34$ and $\delta=1/3$ is assumed, while on the right $\gamma_{\rm inj}=2.07$ and $\delta=0.6$. The symbols represent the data from different experiments as specified in Ref.~\protect\refcite{12bacompo}\protect; the different curves correspond to different realisations of the source distributions and the staircase like curve represents the average over the different realisations.}
\label{fig:anis}
\end{figure}

One thing that we can try to do in order to obtain more refined constraints on the sources from CR observations at Earth is to go beyond the simplifications of the leaky box description of CR propagation. In particular, one ingredient that is missing in the leaky box model and in most calculations of the CR propagation from their sources to Earth (through e.g. GALPROP\cite{galprop} or DRAGON\cite{dragon}) is the discrete nature of these sources in space and time. Calculations taking this effect in account have only recently appeared in the literature\cite{ptuseo,12bacompo,12baanis} and show two important facts, that Figs.~\ref{fig:allpart} and \ref{fig:anis} should help illustrate.

In the calculations behind the plots in these figures, CRs are assumed to be injected by SNRs that explode at random places in the Galaxy, with a probability following the spatial distribution of pulsars and with a rate ${\cal R}_{\rm SN}=(1-3)/100$ yr$^{-1}$. Each SNR is assumed to accelerate particles with an efficiency ranging between 3-15\%, and the accelerated particles then propagate through the Galaxy and to Earth with a rigidity dependent, spatially homogeneous diffusion coefficient, that is normalised in such a way as to ensure that the B/C ratio can be reproduced (see Ref.~\protect\refcite{12bacompo}\protect for details).
Fig.~\ref{fig:allpart} shows the all particle spectrum resulting from two different description of injection and propagation. In the left panel we compare with data the spectrum that is obtained assuming $\gamma_{\rm inj}=2.34$ and $\delta=1/3$, while the right panel corresponds to $\gamma_{\rm inj}=2.07$ and $\delta=0.6$. It is apparent that the data can be reproduced reasonably well in both models up to the energies corresponding to the transition between Galactic and Extragalactic CRs. The situation changes radically, however, when one considers the anisotropy expected in these two different scenarios. This is shown in Fig.~\ref{fig:anis}, where the left and right panel corresponds to the same situations considered in the left and right panel of Fig.~\ref{fig:allpart}.

In a regime of diffusive propagation, the anisotropy is defined as 
\begin{equation}
\vec \delta_A=-\frac{3 D(E)}{c} \frac{\vec \nabla n_{\rm CR}}{n_{\rm CR}}
\label{eq:anis}
\end{equation}
and hence it is especially sensitive to the diffusion coefficient. Indeed, the degeneracy between the two different propagation scenarios is broken in Fig.~\ref{fig:anis}, which clearly shows that for flat source spectra and fast dependence of the diffusion coefficient on energy the amplitude of the anisotropy is systematically overestimated by a large factor.

One could object that actually, based on Fig.~\ref{fig:anis}, also the scenario on the right has problems at accounting for the observations, in spite of the fact that the discrepancy is less than on the left. However this latter discrepancy can be explained and overcome if the actual distribution of nearby sources and the subtleties involved in the measurements are taken into account\cite{sveshni}, leading to the conclusion that the ensemble of different observables favours a scenario in which SNRs inject a relatively steep particle spectrum $\gamma_{\rm inj}\approx 2.3-2.4$ and propagation in the Galaxy occurs as expected for the case of a Kolmogorov-like spectrum of magnetic turbulence.

\section{Revising NLDSA theory}
\label{sec:nldsarev}
The basic theory of NLDSA predicts a spectrum of CRs in SNRs that is typically flatter than $p^{-4}$. In the cases where $\gamma$-rays are observed and their hadronic interpretation is favoured, the inferred particle spectra are systematically steeper\cite{damspecind} than $p^{-4}$. This apparent discrepancy might actually find an explanation in the subtleties of particle transport in the presence of an amplified magnetic field. Looking back at Eq.~\ref{eq:transp}, one notices the presence of the velocity of the scattering centres $\hat u$ in the places where the fluid velocity usually appears when dealing with test-particle descriptions. The correct velocity to use is actually $\hat u$, which is given by the sum of the fluid velocity and the phase velocity of the scattering waves\cite{ziraptu,damspecind}. Most of the times, however, the latter can be ignored, being coincident with the Alfv\'en speed associated with the unperturbed magnetic field and hence totally negligible with respect to the former. In the presence of magnetic field amplification, however, this may no longer be the case and the wave speed can in principle become relevant. Indeed it was shown that the spectral indices observed in a number of SNRs at GeV and TeV energies can be accommodated within NLDSA with MFA if the wave speed that enters the calculations is taken to be the Alfv\'en speed computed in the amplified field\cite{damspecind}. 

\begin{figure}[h!!!!]
\includegraphics[scale=.24]{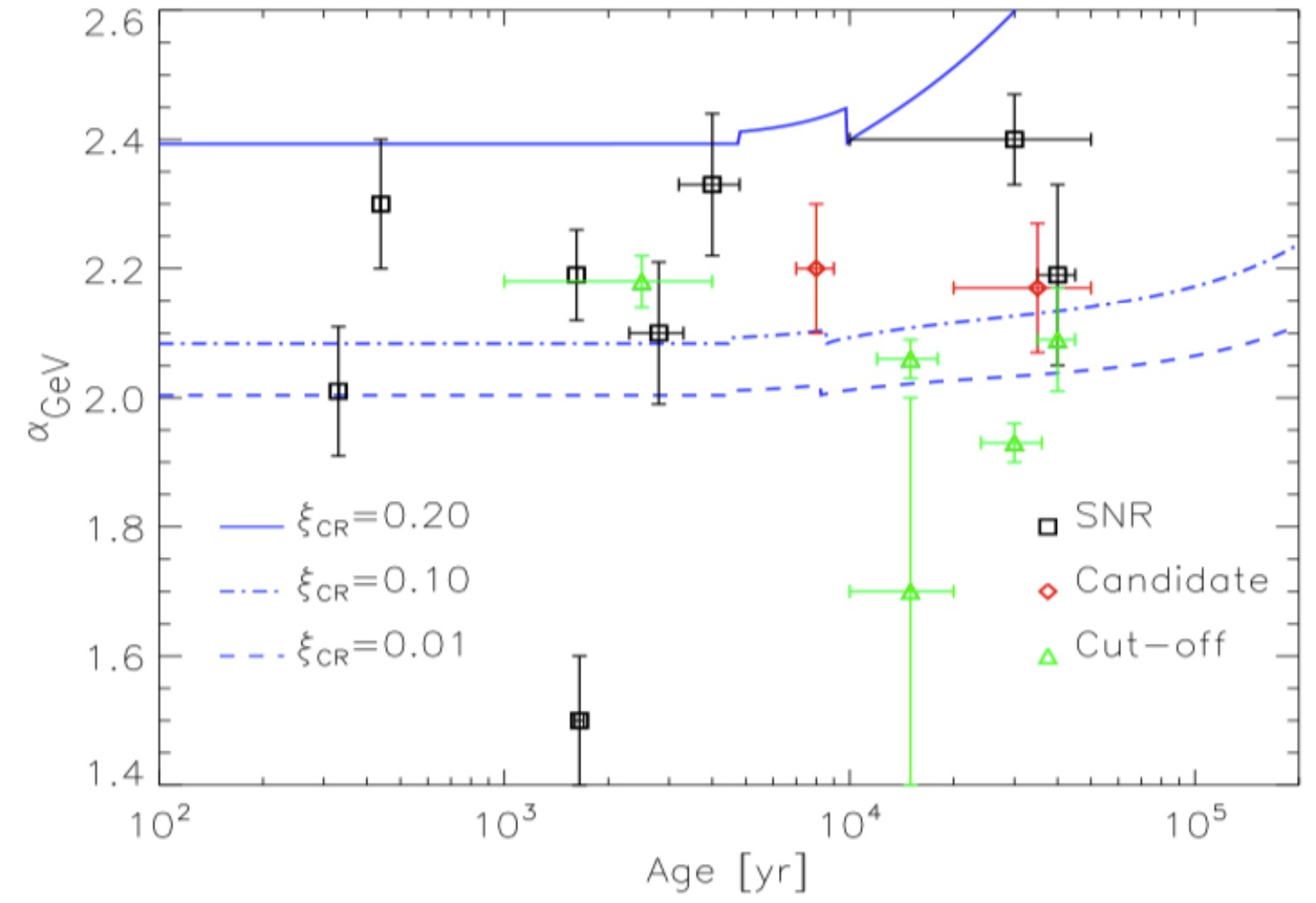}
\includegraphics[scale=.24]{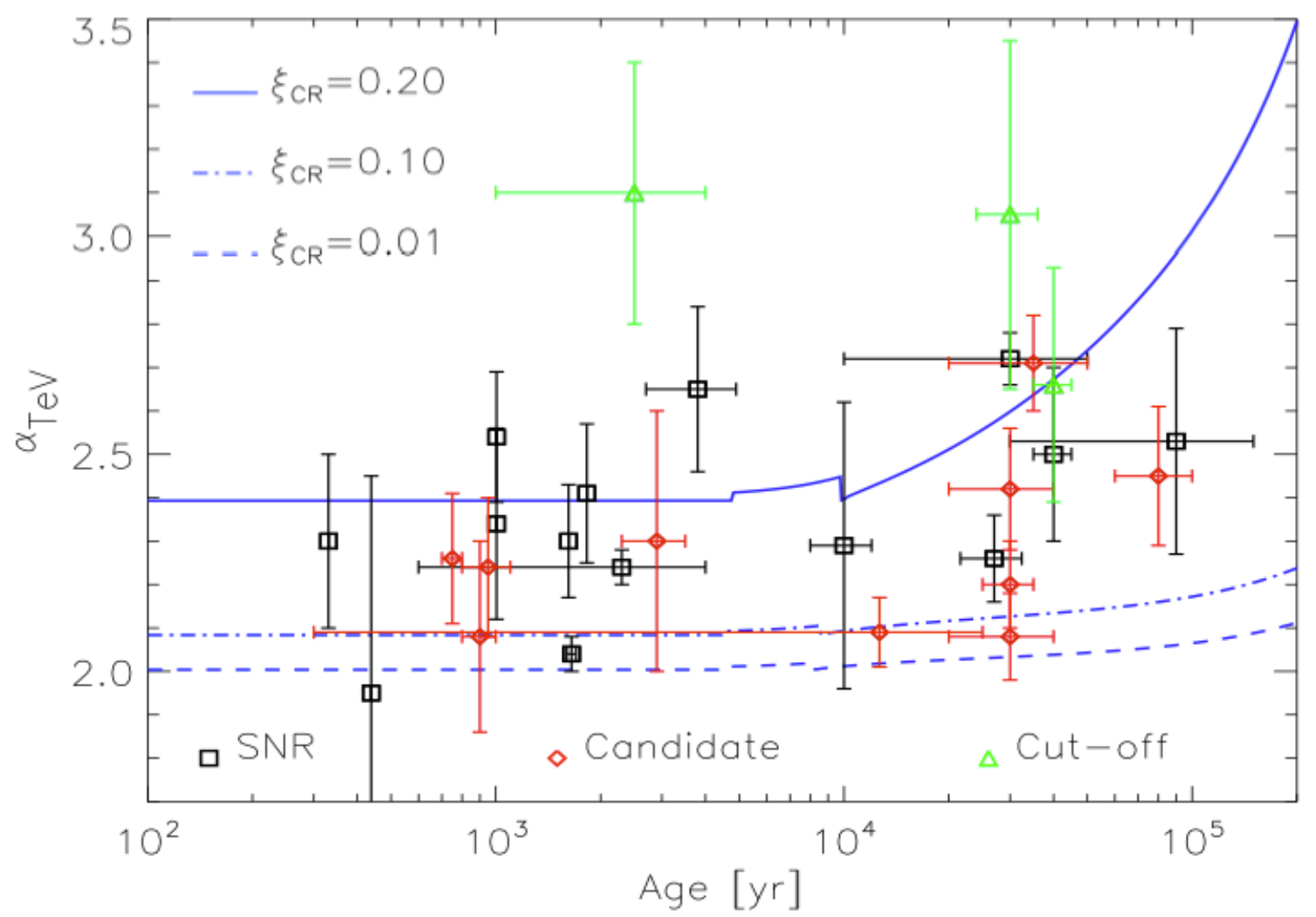}
\caption{Comparison with existing data at GeV (eft panel) and TeV (right panel) energies of the spectral indices predicted for SNRs with different acceleration efficiencies $\xi_{\rm CR}$ (figures taken from Ref.~\protect\refcite{damspecind}\protect). The prediction takes into account the effect of the amplified field in the velocity of the scattering centres.}
\label{fig:damspecind}
\end{figure}
\begin{figure}
\begin{center}
\includegraphics[scale=.4]{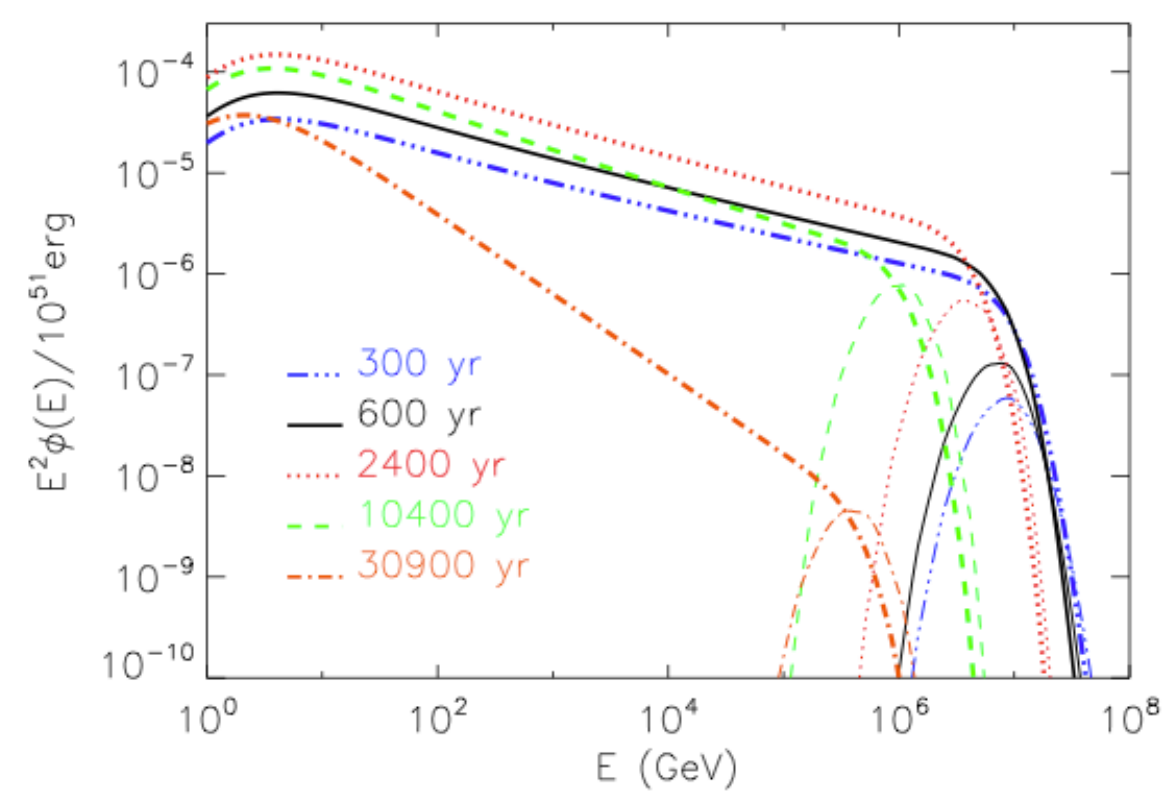}
\caption{Revised prediction for the spectrum released in the ISM by a SNR that is efficiently accelerating CRs (figure taken from Ref.~\protect\refcite{damescspec}\protect). Different line-types refer to different ages, as specified in the figure, while thick lines are for the spectrum advected downstream and thin lines are for the instantaneous escape. The prediction takes into account the effect of the amplified field in the velocity of the scattering centres.}
\end{center}
\label{fig:damescspec}
\end{figure}

When this is done, both the spectra at the sources (see Fig.~\ref{fig:damspecind}) and the spectrum released by the source in the ISM (right panel of Fig.~\ref{fig:damescspec}) can become steeper than $E^{-2}$ even in the presence of efficient acceleration. In fact, the initial theoretical prediction of NLDSA is even overturned: the more efficient the acceleration, the steeper the spectrum.

These results are very interesting but should be taken with caution. The main uncertainty is related to the propagation direction of the magnetic perturbations, which leads to a corresponding uncertainty on whether the modified Alfv\'en speed should be added or subtracted from the plasma motion on the two sides of the shock. In principle, magnetic field amplification could even lead to flatter spectra, a possibility also invoked in the literature already in early studies of shock acceleration physics\cite{blandeich}. 

\section{Revising MFA theory}
\label{sec:mfanew}
While the modification of the NLDSA theory just described was prompted by the need of explaining new data, another important revision of the framework described in Sec.~\ref{sec:nldsa} is imposed by the discovery of a flaw in the equations: the growth rate of the magnetic perturbations in Eq.~\ref{eq:stdrate}, that was adopted in most of the calculations of the last few years, is not the correct one to use in the vicinity of an efficient CR accelerator. Eq.~\ref{eq:stdrate} represents an excellent approximation to the growth rate of waves only in the so-called weakly driven regime, namely when CRs are few and their streaming velocity is not very large, in other words when the current they carry is not very large.

Let us consider a system that is made of a positively charged cloud of CRs propagating in the ISM at the shock speed, $v_S$, and let us assume that the background magnetic field of strength $B_0$ is parallel to the shock normal (namely parallel to $v_s$. In the shock frame CRs, with density $n_{\rm CR}$, are isotropic and have a power-law spectrum, while the ISM appears as a cold quasi-neutral plasma, with ions, whose density we call $n_i$, drifting at $v_S$, and electrons with density $n_e=n_i+n_{\rm CR}=n_i(1+\lambda_{\rm CR})$ moving at a slightly lower speed $v_e=v_S/(1+\lambda_{\rm CR})$, so as to ensure charge and current neutrality.

The dispersion relation for parallel propagating waves ($\vec k \parallel \vec B_0$) in such a system is found as\cite{krall}:
\be
\frac{c^2 k^2}{\omega^2}=1+\sum_s \frac{4 \pi^2 q_s^2}{\omega}\int dp \int d \mu \frac{p^2 v(p) (1-\mu^2)}{\omega-k v \mu \pm \Omega_s}\left[\frac{\partial f_{0,s}}{\partial p} + \frac{1}{p} \left(\frac{v k}{\omega}-\mu\right) \frac{\partial f_{0,s}}{\partial \mu}\right]
\label{eq:gendisp}
\ee
where the sum is over the plasma species $s$, of electric charge $q_s$, and $f_{0,s}$ is the unperturbed distribution function of each species. The underlying assumption is that the $f_{0,s}$ are gyrotropic, namely only dependent on the modulus of $\vec p$ and on the pitch angle $\mu$. In addition $\Omega_s=q_s B_0/m_s \gamma_s c$ is the relativistic gyrofrequency of each species, with $\gamma_s$ the appropriate Lorentz factor.
Adopting a spectrum $f(p)\propto p^{-4}$ for the accelerated particles, and considering only low frequency waves $\omega\ll k v_S$, the dispersion relation that one finds reads\cite{achterberg83,zweibel03,bell04,ab09}:
\be
\omega^2=v_A^2 k^2+\lambda_{\rm CR} k v_S \Omega_{ci}^* \left[\Phi(x)+i \Psi(x)\right]
\label{eq:findisp}
\ee
with 
\be
\Phi(x)=\pm \frac{1}{2} \left\{1-\frac{[1-x^2]}{2x} \ln \left|\frac{1+x}{1- x}\right|\right\}\, \, \, \,  {\rm and} \, \, \, \, \, \, \, \, \Psi(x)=\frac{\pi}{4}
\left\{
\begin{array}{ccc}
x &  & x<1 \\
1/x &  & x>1 \\
\end{array}
\right.
\label{eq:phi}
\ee
In the above expressions $x=k r_{L,0}$, with $r_{L,0}$ the Larmor radius of the lowest energy particles being accelerated, and $\Omega_{ci}^*=e B_0/m_p\ c$ is the non relativistic ion cyclotron frequency. 

The above dispersion relation is found to have rather different solutions in the regime of strong and weak CR current, namely depending on whether
\be
\frac{4 \pi}{c} J_{\rm CR}\, _<^>\, k B_0\ .
\label{eq:critcurr}
\ee
For weak CR currents an excellent approximation for the solution of Eq.~\ref{eq:findisp} is provided by:
\be
{\cal R}e(\omega)=k v_A, \, \, \, \, {\cal I}m (\omega)=\frac{\pi}{8} \frac{v_S}{v_A} \frac{n_{\rm CR}}{n_i} \Omega_{ci}^* \left\{
\begin{array}{ccc}
x &  & x<1 \\
1/x &  & x>1\ . \\
\end{array}
\right.
\label{eq:weakdisp}
\ee
Eq.~\ref{eq:weakdisp} essentially states that the perturbations that grow effectively are Alfv\'en waves and they only grow effectively in a range of wavelengths such that there are particles able to produce them resonantly: this condition is realised for all $k$ such that $k<1/r_{L,0}$ and the wave-growth peaks at $k=1/r_{L,0}$, which is where the number of resonant particles is largest. An important thing to notice is that ${\cal I}m(\omega)$ in Eq.~\ref{eq:weakdisp} is essentially the same as $\Gamma_{\rm res}$ in Eq.~\ref{eq:stdrate}, and this is also the growth rate that was used for CR induced waves not only in the works dealing with CR propagation in the Galaxy, for which situation the weak current approximation is appropriate, but also, erroneously, for wave growth in the vicinity of a CR modified shock (see {\it e.g.} Ref.~\protect\refcite{ab06}\protect), where the CR current is certainly not weak. Indeed, the condition in Eq.~\ref{eq:critcurr} can be rewritten in terms of CR acceleration efficiency, obtaining that the current is strong as soon as:
\be
\xi_{\rm CR}\, _\sim^>\, \frac{\Lambda}{3} \frac{v_A^2 c}{v_S^3}\approx 10^{-4} \left(\frac{B_0}{\mu G}\right)^2 \left(\frac{v_S}{10^9 {\rm cm}{\rm s}^{-1}}\right)^{-3}\ ,
\label{eq:criteff}
\ee
where 
\be
\xi_{\rm CR}=\frac{P_{\rm CR}}{n_i\ m v_S^2}=\frac{1}{n_i m v_S^2}\int_{p_{\rm min}}^{p_{\rm max}}dp p^2\ \frac{vp}{3} f(p)
\label{eq:xicr}
\ee
 is the fraction of kinetic energy of the blast wave that gets converted into CRs, and we took $\Lambda=\ln(E_{\rm max}/E_{\rm min})=15$.

For the case of strong current, the solution of Eq.~\ref{eq:findisp} for the resonant modes ($x<1$) can be written as:
\be
{\cal R}e(\omega)={\cal I}m(\omega)=\sqrt{\frac{\pi}{8} \frac{n_{\rm CR}}{n_i} \frac{m v_S}{p_{\rm min}}} \Omega_{ci}^* x\ .
\label{eq:strongdisp}
\ee 
For large wave-numbers, where the instability is excited non-resonantly, the solution depends on the polarisation. 

\begin{figure}
\includegraphics[scale=.6]{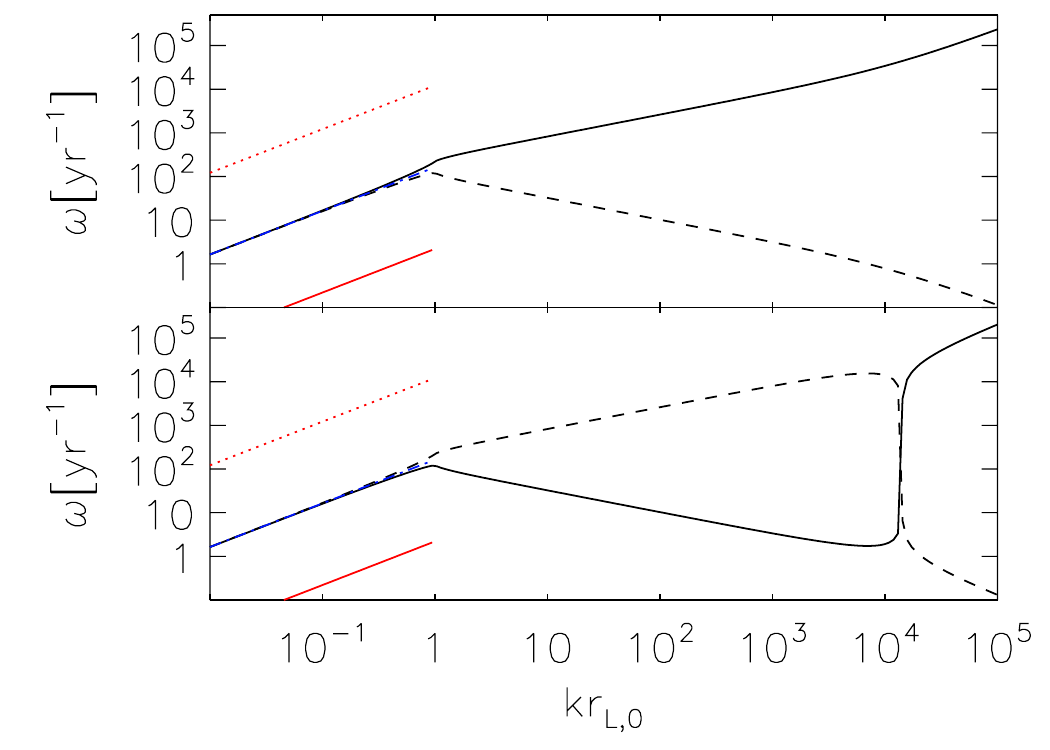}
\caption{Dispersion relation for parallel waves induced by the cosmic ray streaming in a magnetised quasi-neutral plasma. The solid and dashed curves represent the real and imaginary part of the frequency respectively. The upper (lower) panel refers to the left (right) hand polarised mode. The dot-dashed (blue) curve represents the approximation in Eq.~\ref{eq:strongdisp}, while the dotted (red) curve represents the growth rate classically assumed for the streaming instability (Eq.~\ref{eq:stdrate} and Eq.~\ref{eq:weakdisp}), appropriate to describe the growth in a regime of weak current (see discussion in the text).} 
\label{fig:disprel}
\end{figure}

The general solution of the dispersion relation written in Eq.~\ref{eq:findisp} is plotted in Fig.~\ref{fig:disprel} for values of the parameters typical of a SNR shock that is efficiently accelerating CRs, namely: $v_S=10^9{\rm cm}{\rm s}^{-1}$, $\epsilon_{\rm CR}=0.1$. Together with the solution of the dispersion relation (solid lines for the real part of $\omega$ and dashed lines for its imaginary part), also the approximate expressions for the growth-rate of the resonant branch in Eq.~\ref{eq:weakdisp} and Eq.~\ref{eq:strongdisp} are shown as a dotted (red) curve and as a dot-dashed (blue) curve respectively. It is clear that adopting the expression in Eq.~\ref{eq:weakdisp} leads to overestimate the growth-rate of the instability by a large factor.

From the physics point of view, what had been neglected, when blindly taking the growth rate of the streaming instability from studies devoted to CR propagation in the ISM, was the fact that when CRs are numerous and fast, they deeply affect the dispersion relation of parallel propagating waves, changing not only the imaginary par of the frequency, but also the real part. The end result is that modes that are resonant with the streaming particles grow more slowly (black dashed line versus red dashed line in both panels of Fig.~\ref{fig:disprel}) but at larger frequencies (black solid line versus red solid line). This latter fact is potentially very interesting, because the phase speed of the waves ${\cal R}e(\omega)/k$ is in this case much larger than $v_A$. We already discussed in the previous section the effects that the velocity of the scattering centres might have on the particle spectra when it becomes a non-negligible fraction of the fluid velocity. For typical values of the parameters, from Eq.~\ref{eq:weakdisp} we find:
\be
v_\phi=\frac{{\cal R}e(\omega)}{k}=\sqrt{\frac{3\pi}{8 \Lambda}\frac{v_S}{c} \xi_{\rm CR}}v_S\approx 2\times 10^7 {\rm cm\ s}^{-1} \left(\frac{v_S}{10^9 {\rm cm\ s}^{-1}}\right)^3\ . 
\label{eq:phasespeed}
\ee 
This speed is $\approx 2 \%$ of the shock speed for fast shocks ($v_S \approx (0.5-1)\times 10^9 {\rm cm\ s}^{-1}$) and can easily lead to a 10-15 \% change of the spectral index, maybe more, depending on what happens to the propagation of the waves in the downstream. 

Another important feature to notice in Fig.~\ref{fig:disprel} is the fact that in this regime of strong current the modes with different polarisation grow differently. In particular, while the growth of the left-hand polarised mode (the one that rotates in the same sense as positive charges in $\vec B_0$) quickly drops at large $k$-vectors, where there are no particles to excite waves resonantly, the right-hand polarised mode grows efficiently especially at large wave-numbers. Indeed for $k>1/r_{L,0}$, the perturbations become quasi-standing oscillations (${\cal R}e(\omega)<<{\cal I}m(\omega)$) and their growth is maximum at 
\be
k_{\rm crit}=4 \pi J_{\rm CR}/(cB_0), 
\label{eq:kcrit}
\ee
where
\be
{\cal I}m(\omega(k_{\rm crit}))=v_A k_{\rm crit}\ .
\label{eq:gcrit}
\ee

This non-resonant short wavelength mode is usually referred to as ``Bell's mode'', since it was first highlighted by Bell\cite{bell04} in 2004. Its very large growth, promising to lead to amplified field in the range of few $100 \mu G$, has made it the subject of much attention in the past ten years. However, at least in principle, this mode, being very short wavelength (the larger the CR current, the shorter the maximum growing wavelength, see Eq.~\ref{eq:kcrit}) is not the most promising to guarantee efficient particle scattering, and hence to lead to particle acceleration up to high energies. On the contrary, in the short wavelength turbulence that one expects as the result of Bell's mode, the scattering will be rather inefficient, especially at high energies, since a scaling of the diffusion with $p^2$ rather than $p$ is expected, with $p$ the particle momentum.
In order to understand, then, the appeal of Bell's mode, let us consider the level of magnetic field amplification that is expected based on the growth rates found above.

The evolution of magnetic power in a CR induced precursor can be written as:
\be
u\frac{\partial {\cal F}}{\partial x}=\sigma {\cal F}\ ,
\label{eq:dfdx}
\ee
where $\sigma=2 {\cal I}m(\omega)$ and ${\cal F}$ is the normalised magnetic energy density per unit logarithmic band width.
The magnetic power at the shock can be estimated analytically in the test particle case, where the fluid velocity in the upstream can be considered constant ($u=v_S$) and the particle spectrum is $f(p)\propto p^{-4}$. Let us consider only resonant modes. With the growth rate in Eq.~\ref{eq:weakdisp}, we can write
\be
\sigma_w=\frac{\pi}{4} \Omega_{ci}^* \frac{v_S}{v_A} \frac{n_{\rm CR}^{\rm res}}{n_i}\ .
\label{eq:sigmaweak}
\ee
Our goal is that of expressing the power in amplified magnetic field as a fraction of the total power converted into CR acceleration. In order to do so, we express $n_{\rm CR}^{\rm res}$ in the above equation in terms of the CR distribution function
\be
n_{\rm CR}^{\rm res}=4 \pi \frac{p^4 f(p)}{p_{\rm res}}=\frac{4 \pi D}{v_S}\frac{p^4} {p_{\rm res}}\frac{\partial f}{\partial x}
\label{eq:nresf}
\ee
where in the last equality we have used the steady-state transport equation integrated between upstream infinity and the shock. If we then remember the general expression for the parallel diffusion coefficient as a function of ${\cal F}$ (Eq.~\ref{eq:qldiff}), $D(p)=4v(p)r_L(p)/{3 \pi \cal F}$, we find:
\be
\sigma_w=\frac{8\pi}{3} \frac{8 \pi v_A}{B_0^2 {\cal F}} \left(p^4 v(p)\frac{\partial f}{\partial x}\right)_{\rm res}\ ,
\label{eq:sigmaw}
\ee

Using this expression in Eq.~\ref{eq:dfdx} one finds:
\be
\frac{\partial {\cal F}}{\partial x}=\frac{8 \pi}{3} \frac{v_A}{v_S} \frac{8 \pi}{B_0^2}\left(p^4 v \frac{\partial f}{\partial x}\right)_{\rm res}
\label{eq:dfdxfin}
\ee
which is easily integrated in $x$ to lead to:
\be
{\cal F}_0=\frac{2}{\Lambda} \frac{v_S}{v_A} \frac{P_{\rm CR}}{n_i m v_S^2}
\label{eq:F0std}
\ee
and in terms of magnetic field at the shock, by use of Eq.~\ref{eq:dbf}:
\be
\left(\frac{\delta B}{B_0}\right)=\sqrt{2 \frac{v_S}{v_A} \xi_{\rm CR}}\ \approx 30 \left(\frac{v_S}{10^9 {\rm cm\ s}^{-1}}\right)^{1/2}\left(\frac{\xi_{\rm CR}}{0.1}\right)^{1/2},
\label{eq:dbweak}
\ee
where $P_{\rm CR}$ and $\xi_{\rm CR}$ are as defined in Eq.~\ref{eq:xicr} with $f$ taken at the shock position.

The same kind of calculations, when carried out with the modified growth rate in Eq~\ref{eq:weakdisp}, appropriate to describe the strongly current driven regime, lead to:
\be
\left(\frac{\delta B}{B_0}\right)=\left({\frac{8 \Lambda }{3 \pi} \frac{c}{v_S} \xi_{CR}}\right)^{1/4}\approx 2.5 \left(\frac{v_S}{10^9 {\rm cm\ s}^{-1}}\right)^{-1/4}\ \left(\frac{\xi_{\rm CR}}{0.1}\right)^{1/4} .
\label{eq:dbstrong}
\ee

The fact that the amplification factor is inversely proportional to the shock speed might suggest that in the strongly current driven regime, this instability is more effective at slow shocks, but in fact the maximum amplification factor is easily estimated by using the condition that, in order for this description to hold, the upper inequality in Eq.~\ref{eq:critcurr} must be satisfied at $k=1/r_{L,0}$. When this is done, what one finds is that the maximum achievable resonant amplification of the field at a shock that is efficiently accelerating CRs is:
\be
\left(\frac{\delta B}{B_0}\right)_{\rm MAX}=\left[\frac{8}{\pi} \left(\frac{\Lambda c}{3 v_A}\right)^{2/3}\right]^{1/4} \xi_{\rm CR}^{1/3}\approx 5.5 \left(\frac{\xi_{\rm CR}}{0.1}\right)^{1/3}\ .
\label{eq:dbmaxres}
\ee
and this would be realised for shock speeds:
\be
v_S\approx \left(\frac{\Lambda}{3}\frac{v_A^2c}{\xi_{\rm CR}}\right)^{1/3}\approx 4 \times 10^7 {\rm cm\ s}^{-1}\left(\frac{\xi_{\rm CR}}{0.1}\right)^{-1}\ .
\label{eq:bestspeed}
\ee
These shocks are not expected to be accelerating CRs effectively.

The conclusion that one is forced to draw at the end of this session is that the vast amount of work done on the self-consistent inclusion of magnetic field amplification and back reaction into the description of CR modified shocks needs extensive revisions. The picture in which the high magnetic fields derived from observations of SNRs\cite{apjl08} and the acceleration of particles up to the {\it knee} energy\cite{bac07} could both be explained by means of the resonant streaming instability alone was internally flawed, by the assumption of a growth rate of the instability that was not appropriate for the regime of efficient acceleration.

As far as magnetic fields are concerned, if large amplification factors are to be achieved, the resonant streaming instability is not a viable mechanism, while the non-resonant mode of the same instability could do the job. The growth at $k_{\rm crit}$ of the right hand polarised mode is extremely fast in the linear phase and leads to expect extremely high amplification factors. The question of what is the mechanism by which the instability saturates has been the subject of extensive numerical study, both through MHD\cite{bell05,zira08} and PIC\cite{niemec08,riquspit09,ohira09} simulations, and the result is that even the most conservative estimates give fields that easily reach $30-100 \mu$ G strengths for fast shocks and CR acceleration efficiency $\xi_{\rm CR}\approx 0.1$. 
Another important result of these numerical investigations is the fact that the wavelength at which the field growth is the fastest ($1/k_{\rm crit}$) progressively increases with increasing field strength, scaling with $(\delta B/B_0)^2$. This fact is of fundamental importance for particle acceleration. Indeed, one of the most serious objections against the non-resonant instability as the explanation for reaching high energy is the fact that the wave modes that grow are too high frequency to induce efficient particle scattering. However recent developments seem to support the contrary. In the picture in which the instability is induced by the entire population of CRs propagating in the ISM with the shock speed, an inverse cascade by almost 10 orders of magnitude would be needed to efficiently scatter particles with energy close to the {\it knee} (from a scale that is of order $10^{-3}-10^{-2}\ r_{L,0}$ up to the gyro radius of PeV protons). This is very hard to imagine, but two facts might help. First of all, the turbulence might be seeded far from the shock by the particles escaping from the system at the maximum energy: this would reduce $k_{\rm crit}$ by a factor $p_{\rm min}/p_{\rm esc}$ (see Eq.~\ref{eq:kcrit} and consider that the current carried by the escaping particles for a $p^{-4}$ spectrum is $J_{\rm esc}=J_{\rm CR}(p_{\rm min}/p_{\rm esc})$) and still lead to a fast enough growth of the perturbations. In addition, the fact that Bell's turbulence has a well defined helicity (only right-hand polarised modes grow efficiently) creates the basis for a dynamo mechanism, that then effectively moves power to larger scales\cite{bykov11,bykov13}. 

A very recent development is the appearance of numerical studies that include particle acceleration and magnetic field amplification in a self-consistent way, rather than considering the development of the instability induced by a fixed current, as was always the case in the past\cite{damtolik13,damtolik14}. These studies indeed confirm that in the strongly current driven regime, the escaping particles seed the growth of small wavelength perturbations far upstream, but close to the the shock there is considerable power at the resonant scales. 

This picture in which the escape of particles is the main source of instability and Bell's modes are the main source of scattering has recently led to propose a paradigm shift: while in the standard picture of CR acceleration in SNRs the highest energy particles were assumed to be accelerated in the early stages of the Sedov-Taylor phase of expansion, a very recent proposal\cite{schurebell1,schurebell2} wants these particles accelerated at a much earlier time, in the first few years after the explosion of a type II Supernova expanding in the dense wind of its progenitor star. According to this paradigm Cas A could have been a PeVatron once and signatures of that time could still be detectable with upcoming $\gamma$-ray instruments.

\section{News from the optical}
\label{sec:neutrals}
In the last few years a very interesting development in the quest for a final proof of the CR-SNR association has come from the most traditional kind of observations, namely observations in the optical band. It is well known that several SNRs show Balmer dominated shocks\cite{chevalierray78}. The Balmer $H_\alpha$ emission is produced when neutral hydrogen de-excites after being excited by collisions with the hot plasma behind the shock. When a collisionless shock propagates in a partially ionised medium, the neutral particles are not affected by the shock and therefore they acquire a velocity and temperature difference with respect to the ionised component, which is slowed down and heated. 
Due to this kinematic difference, behind the shock, the processes of charge exchange and ionisation start to take place in a non-equilibrium situation: this leads to Balmer emission, with a line that is usually made of two components, a narrow component, associated with emission by cold neutrals that have kept their upstream distribution, and a broad component, due to neutrals that result from a charge-exchange reaction in the downstream (hot ions that have become neutrals) and have an effective temperature close to that of the downstream ions. 

In the presence of efficient CR acceleration, as we mentioned in Sec.~\ref{sec:nldsa}, one expects the downstream plasma temperature to be lower than the shock jump conditions would predict, and hence the broad Balmer line should become narrower than expected. At the same time, if a CR precursor develops (such as described in Sec.~\ref{sec:nldsa}), a velocity and temperature difference between ions and neutrals will come about even before the shock transition. In this case charge exchange processes will take place out of equilibrium also in the precursor, producing some heating and slowing down of the neutrals already in the upstream. This might result in a broader than expected narrow line (see Ref.~\protect\refcite{heng10}\protect for a review).
  
A great deal of excitement was aroused in recent years from the measurement of an anomalous shape of the Balmer line in a couple of SNRs\cite{helder09,helder10}. The authors of the discovery interpreted their findings in terms of efficient CR acceleration at these shocks, deducing $\xi_{\rm CR}>50\%$ for the case of SNR RCW86. However, the physical modelling of the system behind these interpretations was not very accurate, due to the lack of a self-consistent description of a CR accelerating shock propagating in a quasi-neutral medium. Now that such a description has finally become available\cite{chargex1,chargex2,chargex3}, the inferred acceleration efficiencies have revised somewhat, while additional possible signatures of efficient CR acceleration have been highlighted. 

\begin{figure}
\includegraphics[scale=.40]{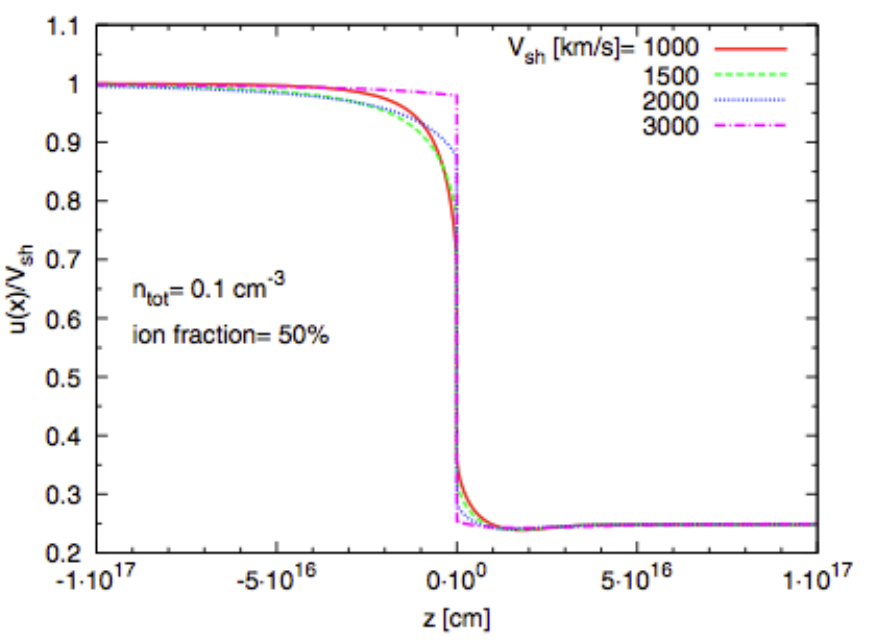}
\includegraphics[scale=.72]{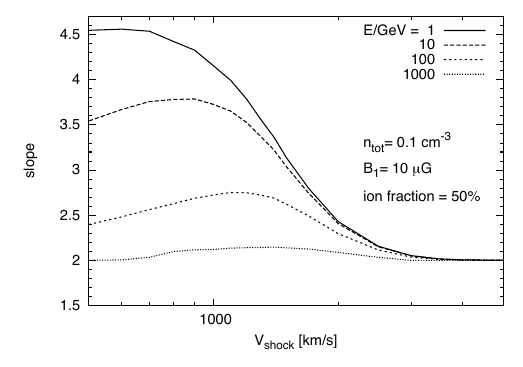}
\caption{Left panel: The neutral induced precursor. The fluid velocity is plotted as a function of distance from the shock (the shock is at z=0 and the upstream is on the left). For velocity lower than 300 ${\rm km\ s}^{-1}$, the fluid slows down already in the upstream. The total compression ratio, however, is never larger than 4. Right panel: slope of the particle distribution function at different energies as a function of the shock speed. The plot refers to the spectrum of test particles accelerated at a shock that propagates in a medium with density $n_{\rm tot}=0.1 {\rm cm}^{-3}$ and $50\%$ ionisation fraction (taken from Ref.~\protect\refcite{chargex1}\protect).}
\label{fig:neutspec}
\end{figure}

The presence of neutrals upstream of the shock, not only provides a diagnostic for the shock, thanks to Balmer emission (which, by the way, is also the only available probe of the ion temperature behind the shock), but, even more important, affects the shock dynamics. The first effect of the presence of neutrals is the creation of a precursor ahead of the shock even in the absence of CR acceleration. This is due to the fact that some neutrals might undergo charge-exchange downstream of the shock with ions that are moving towards the shock. When this happens nothing prevents the newly formed neutrals to recross the shock and end up in the upstream, where they might undergo another charge exchange interaction or get ionised: this phenomenon has been called ``the neutral return flux''. The net result of this sequence of processes is the deposition of energy and momentum in the upstream, and thus the formation of a precursor. This precursor, however, is very different in nature from the one induced by CRs, and the main difference is that, while in the CR precursor the compression factor increases to values larger than the limit value of $R_{\rm tot}=4$, appropriate for a strong shock, in this neutral induced precursor the compression factor is never larger than 4, and depending on the ionisation fraction of the medium and on the shock speed, might become appreciably less than 4 at the shock (see left panel of Fig.~\ref{fig:neutspec}). This difference between the two precursors, which is due to the fact that the shock stays non-radiative in this case (no neutrals can escape the system towards upstream infinity due to ionisation losses) reflects in an important difference between the spectra of accelerated particles: while the CR precursor had the effect of making the spectrum harder, the neutral precursor makes the spectrum steeper. This effect is illustrated in the right panel of Fig.~\ref{fig:neutspec}, where the spectral index $\alpha$ ($\alpha=(p/f)(df/dp)$ where $f(p)$ is the particle distribution function in momentum) is plotted as a function of the shock speed for different particle energies, and for the plasma parameters specified in the figure. 

It is apparent that this fact might help explain the presence of steep particle spectra in sources with shock velocities below 3000 ${\rm km\ s}^{-1}$. 

\begin{figure}
\includegraphics[scale=.44]{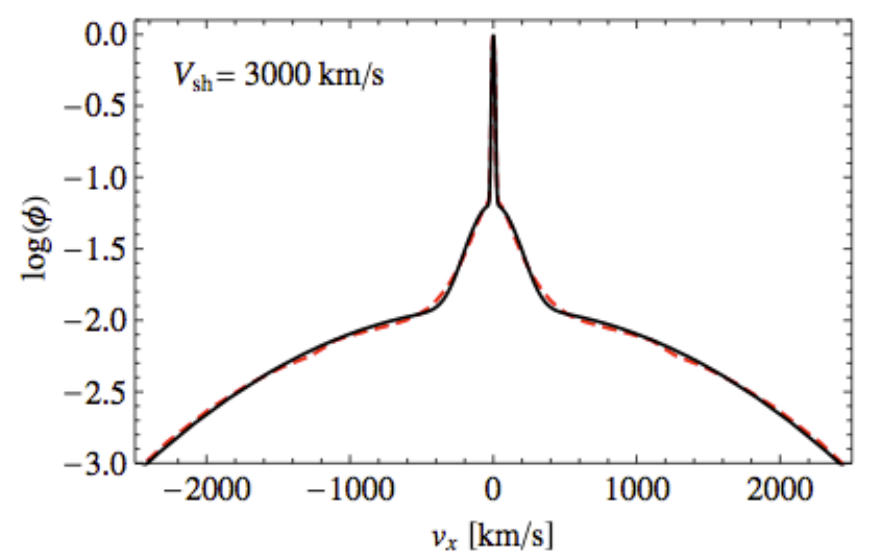}
\includegraphics[scale=.37]{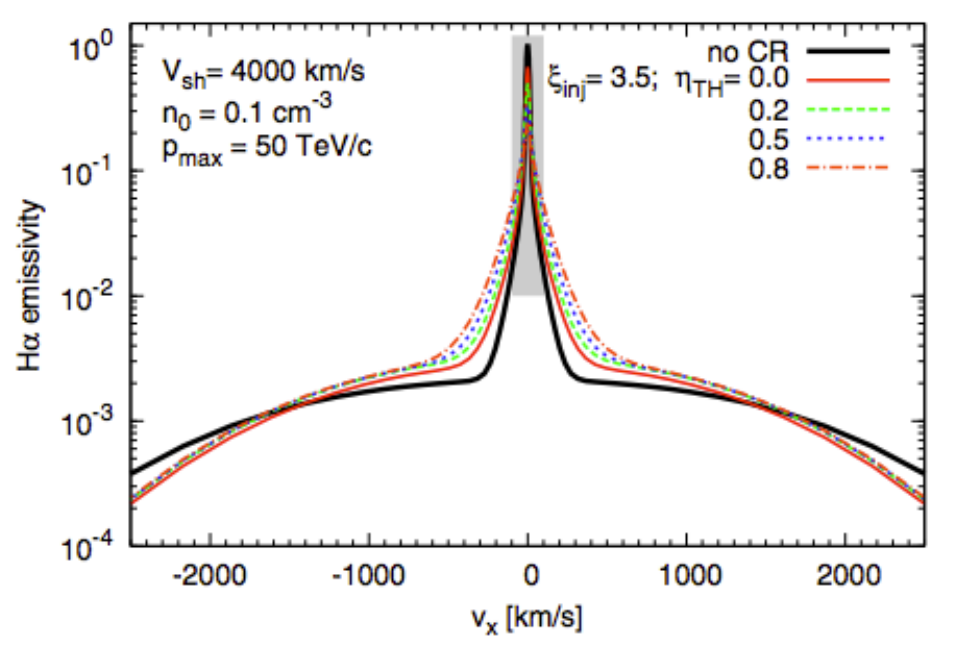}
\caption{Left panel: logarithmic plot of the normalised intensity of the Balmer line, showing the appearance of a third line with intermediate width due to the neutral return flux. The upstream density and ionisation fraction are the same as above and the shock speed is $4000 {\rm km\ s}^{-1}$. Right panel: how the line shape changes in the presence of efficient CR acceleration. A shock with velocity $v_S=4000 {\rm km\ s}^{-1}$, $n_{\rm tot}=0.1 {\rm cm}^{-3}$ and ionisation fraction $50 \%$ is considered. The different lines are as specified in the figure (taken from Ref.~\protect\refcite{chargex2}\protect).}
\label{fig:intermed}
\end{figure}
The presence of a neutral return flux also alters the shape of the Balmer line. Hydrogen atoms that undergo charge-exchange immediately upstream of the shock with ions that have been heated by the neutral return flux give rise to a third component of the Balmer line which is intermediate in width between the broad and the narrow, with a typical width of $\sim 100-300 {\rm km\ s}^{-1}$.
This effect is well illustrated in the left panel of Fig.~\ref{fig:intermed}. Possible evidence for the presence of such an intermediate line might already be present in existing data\cite{ghavamian00}.

\begin{figure}
\includegraphics[scale=0.38]{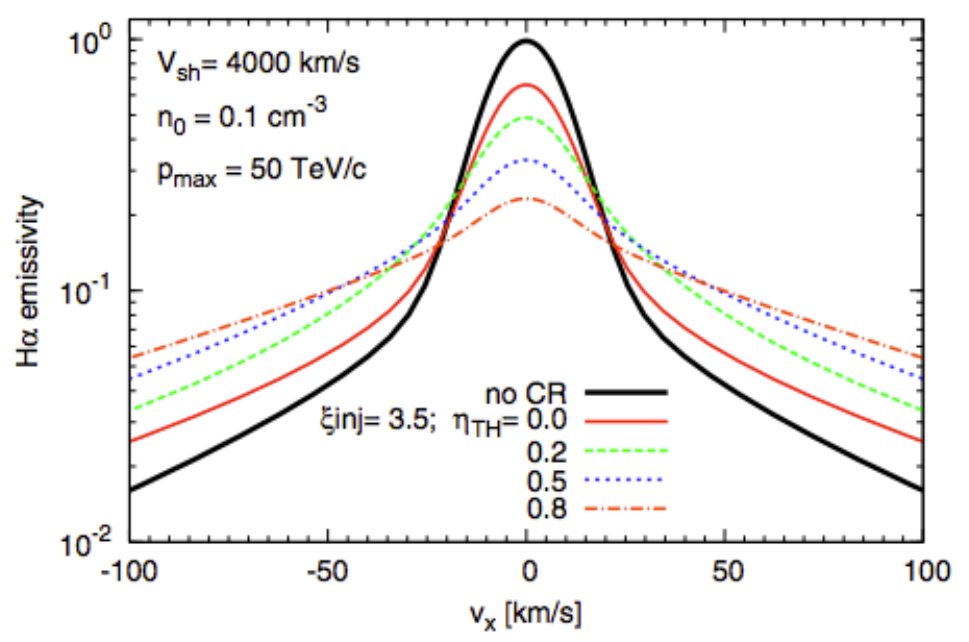}
\includegraphics[scale=0.38]{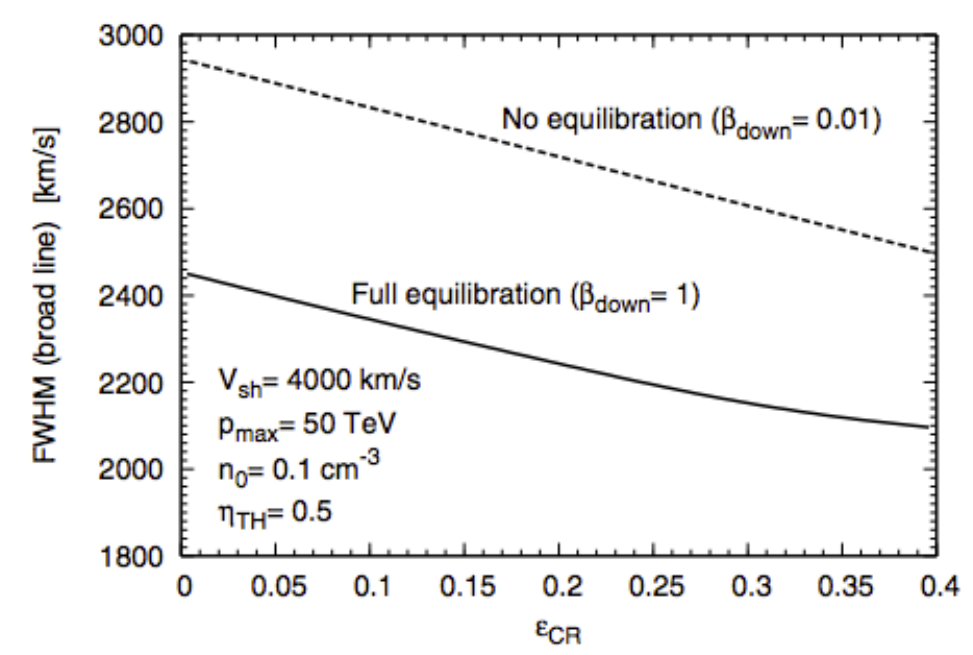}
\caption{Left panel: a closeup of the shaded region in the right panel of Fig.~\ref{fig:intermed} showing in more detail the expected profile of the narrow Balmer line in the absence of CRs and in the presence of efficient CR acceleration ($\xi_{\rm CR}\approx 30\%$) but with different levels of turbulent heating in the precursor. Right panel: the width of the broad Balmer line as a function of CR acceleration efficiency for two extreme assumptions on the level of electron-ion equilibration downstream ($\beta_{\rm down}=T_{e2}/T_{i2}$, the ratio of electron to ion temperature). All the main parameters are specified in the figure (taken from Ref.~\protect\refcite{chargex3}\protect).}
\label{fig:balmerline}
\end{figure}

The effects so far discussed were in the context of a shock that is not accelerating CRs effectively. Let us now move to efficient accelerators, to see how Balmer emission can be used as a diagnostic of CR acceleration efficiency. A complete description of the system can be achieved by solving the fluid equations for the fluid, the transport equation for the accelerated particles and the Boltzman equation for the neutrals, with collisions associated to charge exchange with ions and ionisation. The steady state solution of this set of equations can be found by iteration, and all the thermodynamical quantities, together with the distribution function of neutrals and the spectrum of the accelerated particles can be calculated at each point in space.
In terms of Balmer emission, the presence of CRs qualitatively leads to a broadening of the narrow and intermediate components of the line, and to a narrowing of the broad component (see right panel of Fig.~\ref{fig:intermed}. Quantitatively, however, these effects depend on a number of different parameters. The width of the narrow and intermediate component are especially sensitive to the amount of non-adiabatic heating (turbulent heating) in the precursor, which is a total unknown and can only be parametrised in these calculations. The width of the broad Balmer line, instead, while insensitive to what happens in the precursor, strongly depends on the level of electron-ion temperature equilibration behind the shock.
These effects are illustrated in Fig.~\ref{fig:balmerline}, where we plot, in the left panel, the broadening of the narrow line as a function of turbulent heating in the precursor, and, in the right panel, the width of the broad component as a function of CR acceleration efficiency and for two extreme hypotheses on the ratio of electron to ion temperatures downstream. The strong dependence of the FWHM of the broad line on the ratio of electron to proton temperature is one of the main obstacles to derive firm conclusions on the efficiency of CR acceleration from observations.

This model has been applied to the two SNRs for which reliable measurements of an anomalous shape of the broad Balmer line are available, SNR 0509-67.5\cite{helder10,helder11} and SNR RCW86\cite{helder09}. In the case of SNR 0509-67.5, observations of the SW rim (that hosts a fast shock, with $v_s \approx 5000\ {\rm km\ s}^{-1}$, and is the most promising site to detect efficient CR acceleration) are compatible with an efficiency ranging from 10\% to 50\% depending on $\beta_{\rm down}$ (Ref.~\protect\refcite{morlino0509}\protect). If one, however, takes into account the fact that at such fast shocks the equilibration is usually poor\cite{ghavamianeq}, then the most likely scenario is one in which the CR acceleration efficiency is at the 20-30\% level\cite{morlino0509}. In the case of RCW86, an additional difficulty in constraining the acceleration efficiency comes from the fact that the distance to this remnant is not well known, and this reflects in uncertainties on the shock velocity. Depending on whether the distance is 2 or 3 $kpc$, the measurements are compatible with no acceleration or acceleration with efficiency up to 30\%. However, measurements of the electron temperature are available for this remnant, and when these are combined with Balmer emission, the most likely scenario entails an acceleration efficiency of order 20\%\cite{morlinorcw86}. It might seem striking that in both cases the acceleration efficiency seems to be above the average implied for SNRs as CR factories ($\approx 10\%$). However, it should be kept in mind that these estimates refer, as the measurements of $H_\alpha$ emission do, to small sections of the shock surface. Indeed, recent work using high spatial and spectral resolution observation of Balmer emission\cite{nikolic} has highlighted that SNR shocks can be highly dishomogeneous. 

Further progress is promising to come from detailed modelling of other remnants that show anomalous Balmer emission (e.g. narrow line so broad as to imply a temperature of the upstream plasma incompatible with partial ionization). However, given the complex interplay between the different parameters, in order to achieve reliable conclusions, a combination of optical observations with X-ray observations aimed at measuring the electron temperature is likely what is needed.

\section{News from CR observations from Earth}
\label{sec:directcr}
Direct observations of CRs in recent years have also brought about very intriguing results, concerning both hadronic and leptonic CRs. We briefly illustrate a few of them in the following together with the proposed explanations.

\subsection{Spectral breaks}
\label{sec:hardening}
An important result found by the PAMELA satellite is the hardening of both the proton and He spectra at a rigidity (energy divided by electric charge) of about 230~GV\cite{pamelahard}. In addition, the spectrum of He nuclei is always harder than that of protons: the slope of the proton spectrum below 230~GeV is $\gamma_{1,p}=2.85\pm 0.015$ and changes to $\gamma_{2,p}=2.67\pm 0.03$ above that energy; the slope of He is $\gamma_{1,He}=2.77\pm 0.01$ below 230~GeV/n and changes to $\gamma_{2,He}=2.48\pm 0.06$ above. These results seem to confirm, with a much better statistics, previous findings by CREAM\cite{ahn10} that showed the spectra of proton and He to be different and also that all nuclear species present a spectral hardening at around 200~GeV/n. 

The detection of a hardening around 200~GeV/n for all elements is very intriguing especially because that is about the energy at which according to calculations\cite{wentzel74} dating as far back as the '70s, the regime in which CRs propagate changes: in particular, while at energies lower than about 100~GeV CRs are numerous enough to generate the waves that scatter them, at higher energies, this is not the case, and propagation is expected to occur in turbulence of a different origin and with a different spectrum. 

It is generally believed that the main source of turbulence in the Galaxy are SN explosions, which inject the turbulence at a scale of about 50-100 pc. This turbulence then cascades to smaller scales and a commonly (though not universally) accepted description of the cascading process is provided by Non-Linear-Landau-Damping with coefficients such as to produce a Kolmogorov's spectrum: ${\cal F}(k) \propto k^{-5/3}$, where ${\cal F}(k)$ is defined as in Eq.~\ref{eq:dbf}. At any scale, the evolution of the turbulence will be described by an equation of the form:
\be
\frac{\partial}{\partial k} \left[D_{kk} \frac{\partial W}{\partial k}\right]+\Gamma_{CR} W=q_w(k)
\label{eq:turbgrowth}
\ee
where $D_{kk}\propto v_A k^{7/2} (k {\cal F}(k))^{1/2}$ (see Ref.~\protect\refcite{bas12}\protect), $\Gamma_{\rm CR}$ is given by Eq.~\ref{eq:stdrate} and $q_w(k)$ is a source term in the form of a Dirac $\delta$ at $k_{\rm out}\approx (100 pc)^{-1}$.
An analytical estimate of the transition scale between the dominance of CR induced turbulence and general MHD turbulence in the Galaxy can be obtained by equating the non-linear damping rate $\Gamma_{\rm NL}=D_{kk}/k^2$ to the growth rate of CR induced turbulence $\Gamma_{\rm CR}$.

Using in the expression of $\Gamma_{\rm CR}$ the CR spectrum observed by PAMELA at energies above a few hundred GeV, one obtains that the transition must occur at a scale equal to the Larmor radius of particles with energy:
\be
E_{\rm tr}=228 GeV \left(\frac{R_{d,10}^2 H_3^{-1/3}}{\xi_0.1 E_51 {\cal R}_{30}}\right)^{\frac{3}{2(\gamma_p-4)}}\ B_{0,\mu}^{(2\gamma_p-5)/2(\gamma_p-4)}\ ,
\label{eq:breaken}
\ee
where $R_{d,10}$ and $H_3$ are the radius of the Galactic disk and halo in units of 10 and 3 $kpc$ respectively, $\xi_{0.1}$ is the CR acceleration efficiency per SNR in tens of percent, $E_{51}$ is the SN explosion energy in units of $10^{51}\ {\rm erg}$ and ${\cal R}_{30}$ is the SN rate in units of $1/(30 {\rm yr})$; $B_{0,\mu}$ is the average strength of the Galactic magnetic field in $\mu G$ and finally $\gamma_p$ is the observed particle spectral index (in momentum) at energies above 300 GeV (from PAMELA data $\gamma_p=4.67$). The turbulence spectrum will be dominated by CR induced turbulence at scales smaller than $\lambda\approx r_L(E_{\rm tr})\approx 10^{15}$ cm and by external turbulence at large scales. As a result the diffusion coefficient will change its energy dependence around that energy. At larger energies the scaling will be Kolmogorov's: ${\cal F}(k)\propto k^{-2/3}$, and hence $D(p)\propto E^{1/3}$ (see Eqs.~\ref{eq:dbf} and \ref{eq:qldiff}). At lower energies the solution of the coupled equations for the evolution of turbulence and particle transport gives a stronger energy dependence. As a consequence, for a given slope of the particle spectrum injected by the sources, $\gamma_{\rm inj}$, the diffuse CR spectrum $\propto E^{-\gamma_{\rm inj}-\delta_e}$ (see Eq.~\ref{eq:CRspec}) will be steeper at low energies and then flatten at energies around $E_{\rm tr}$. 

\begin{figure}
\includegraphics[scale=0.36]{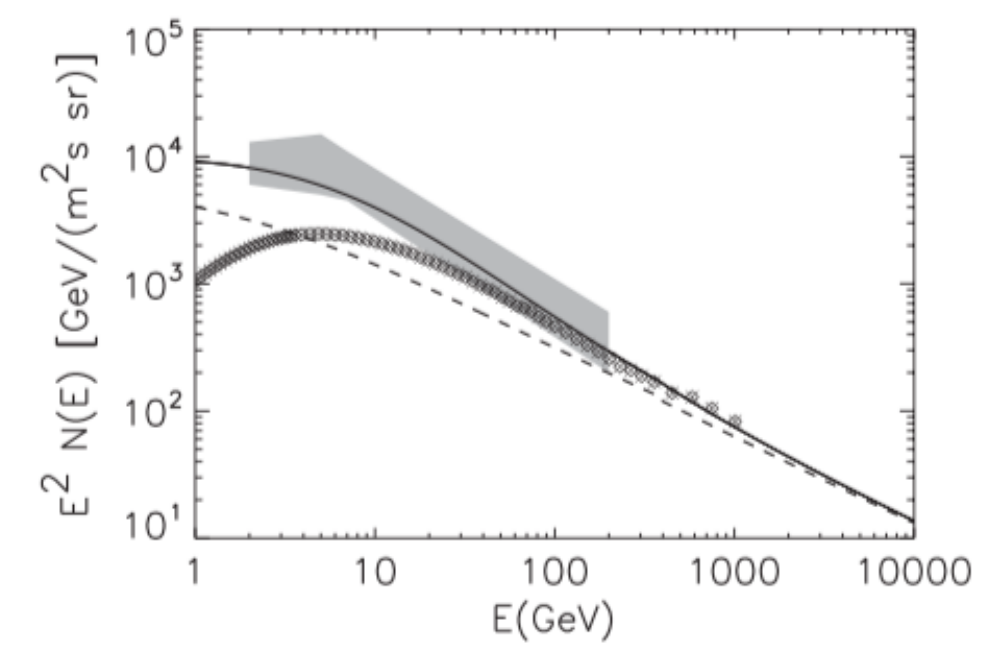}
\includegraphics[scale=0.29]{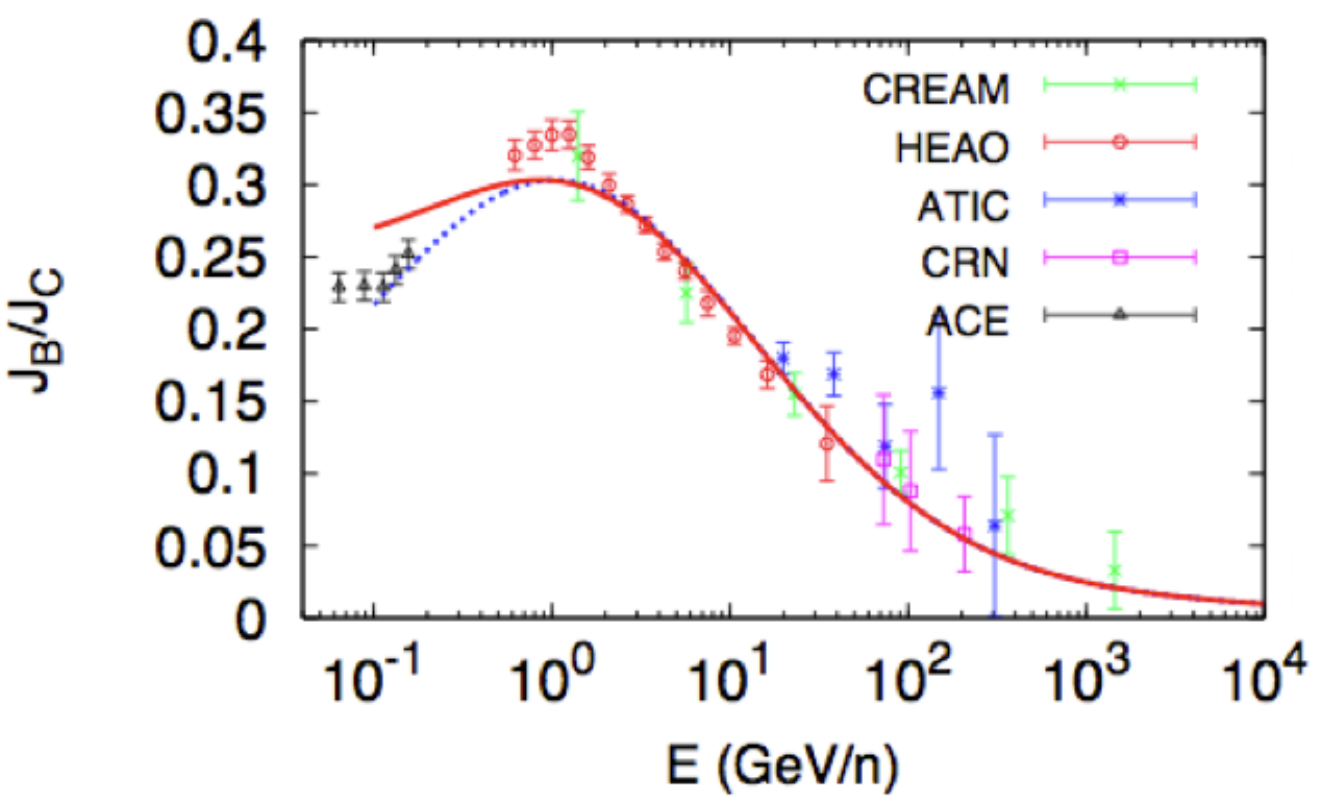}
\caption{Left panel: the proton spectrum computed based on this model compared with PAMELA data (grey circles) and with the CR spectrum deduced from observations of nearby clouds\protect\cite{neronov}\protect (shaded area). Right panel: the predicted $B/C$ ratio as a function of energy per nucleon compared with available data as specified in the figure. }
\label{fig:pamela}
\end{figure}

This very simple model was shown to account reasonably well for PAMELA observations and also for the CR spectrum deduced from observation of nearby clouds\cite{neronov,kachelriess}, as shown in the left panel of Fig.~\ref{fig:pamela}. While in the original version of the model, only the contribution of protons was taken into account for the production of waves, in a follow up work\cite{aloisio} all the most abundant nuclear species were taken into account for wave generation and the propagated spectra of both primaries and secondaries were compared with the available data. The right panel of Fig.~\ref{fig:pamela} shows the impressive agreement between the calculated and observed B/C ratio: this is especially important, given the role of $J_B/J_C$ as a primary probe of the energy dependence of the diffusion coefficient, as we already mentioned in Sec.~\ref{sec:prop}.

Two important points must be mentioned before concluding this session. 

First of all, it should be mentioned that the break detected by PAMELA in the proton spectrum, while in agreement with most previous measurements, does not seem to be confirmed by AMS-02. However, at present, the AMS-02 collaboration has not published its results yet, but rather only presented them at the International Cosmic Ray Conference 2013 and published them on their web site\cite{amshard}, so it is difficult to comment on this issue. What can be safely said is that a transition between two different scattering regimes is something that one expected for CRs to occur, and a spectral break pointing it out could also be expected.

Finally, one important remark is that while the mechanism we have been discussing can explain the spectral hardening of the different chemicals, it certainly cannot explain why the spectrum of He should be harder than the spectrum of protons. At present, the only available explanation for this observation is based on preferential injection of He in the acceleration mechanism\cite{malkovhard}. Such an effect might be induced by the fact that particles are more easily injected in the acceleration process if they are not trapped by Alfv\'en waves, and the latter, being mainly produces by protons (the most abundant specie), are less efficient at trapping He.

\subsection{Leptonic CRs}
\label{sec:elpos}
Another extremely interesting result obtained by the PAMELA satellite is the clear detection of a rising positron fraction at energies from a few GeV through 100~GeV\cite{pamelapos}. This result has also been confirmed, with even higher statistics, by AMS-02\cite{amspos}. Let us see why it is so interesting.

The positron fraction is defined as $\chi=\Phi_{e^+}/(\Phi_{e^+}+\Phi_{e^-})$, where $\Phi_{e^+}$ and $\Phi_{e^-}$ are the fluxes of positrons and electrons respectively. While CR electrons can be either primaries (accelerated in the same sources as for protons) or secondaries, CR positrons are secondaries, namely resulting from CR interactions during their propagation through the Galaxy, and associated to the production and decay of charged pions. In this case, by following a reasoning analogous to that illustrated for secondary nuclei in Sec.~\ref{sec:prop} (made only slightly more complicated by the fact that losses are now important), it is possible to demonstrate that the positron fraction must decrease with energy\cite{serpico}. Therefore the fact that it increases is suggestive of the existence of a source of positrons. An intriguing suggestion immediately made after this discovery was that these extra positrons could come from dark matter annihilation in the Galaxy (see Ref.~\protect\refcite{dmrev}\protect for a recent review). In this case, however, one would also expect a contribution in terms of antiprotons and therefore also the ratio of antiproton to proton fluxes ($\Phi_{\bar p}/\Phi_p$) should be affected. This is not observed: $\Phi_{\bar p}/\Phi_p$ does not show any anomaly and perfectly agrees with the expectation of standard propagation theory. 
While a dark matter related origin of the rise of $\chi$ is difficult to rule out completely\cite{serpico}, astrophysical explanations seem favoured at the current time. The most natural place where to look for a source of extra positrons is the magnetosphere of a strongly magnetised, fast spinning neutron star, that may or may not show up as pulsar. Young pulsars (up to several tens of thousands of years ages) are often observed to be surrounded by non-thermal nebulae, Pulsar Wind Nebulae (PWNe in the following). These nebulae arise from the interaction of the pulsar wind with the surrounding medium, either the parent supernova remnant, or, in some cases, the ISM. And it is indeed from observations of PWNe that we have direct evidence that copious production of electron-positron pairs must be occurring in the magnetosphere of the parent pulsar\cite{nicco11}. 

The standard picture of a PWN (see {\it e.g.} Ref.~\protect\refcite{ea14}\protect for a recent review) is indeed as follows. The parent neutron star can be considered a conducting magnetic dipole, whose rotation induces a strong electric field in the surroundings. At the star surface the electric force acting on a charged particle is much stronger than gravity and leads to the extraction of electrons (also protons may potentially be extracted, but the situation there is less clear, due to the poorly known structure of the star crust). These electrons are accelerated to relativistic energies in regions where unscreened electric field parallel to the local magnetic field exists. This causes $\gamma$-ray emission, either through curvature radiation of the accelerating charges or through Inverse Compton scattering on the thermal photons emitted by the star. Photons with energies above 1 MeV in the intense ambient magnetic field are above the threshold for pair production, and a large number of pairs is produced for each primary electron extracted from the star. These pairs are thought to leave the magnetosphere with a Lorentz factor of about 100, and become part of a relativistic outflow that carries most of the rotational energy lost by the star. Confinement of this magnetised pair wind by the surrounding medium gives rise to a shock propagating outward in the confining medium, and to a reverse shock that propagates towards the pulsar up to a distance such that the ram pressure of the outflow is equal to the pressure of the shocked material downstream. At this termination shock particles are accelerated very efficiently and then they form the bright non-thermal nebulae that we observe from the radio through the $\gamma$-ray band, due to synchrotron and Inverse Compton emission. 

While the PWN phenomenology is a very interesting topic in itself (suffice it to say that these are the only sources in which we have direct evidence of PeV particles), for the purpose of the present discussion, what is relevant are the properties of low-energy, radio emitting particles. These have energies in the range between 1 GeV and 1 TeV and a very flat spectrum, with slope between 1 and 1.5. A source of positrons with a flat spectrum is exactly what is needed to reproduce the rising positron fraction observed by PAMELA and AMS-02. 

When wondering whether these pairs can be responsible for the rising positron fraction, the first concern appears to be the mechanism that allows their escape from the cage that confines them in the PWN. Their Larmor radii in the nebular magnetic fields (of strength around a few 100 $\mu G$) are very small and diffusive escape is not effective. However, pulsars are a population of objects with very high proper motion\cite{arzou}, with a peak around a velocity $V_{\rm PSR}\approx 400-500 {\rm km\ s}^{-1}$. Therefore, they are expected to leave the parent remnant at a time $t_{\rm esc}$ that can be estimated as $V_{\rm PSR} t_{\rm esc} = R_{\rm ST} (t_{\rm esc}/t_{\rm ST})^{2/5}$, assuming that the remnant is expanding in the Sedov-Taylor phase (see Sec.~\ref{sec:acc}). For pulsars in the peak of the velocity distribution, one obtains $t_{\rm esc}\approx 40-50\ kyr$. Indeed, pulsars with ages of tens of $kyr$ are observed to form Pulsar Bow Shock Nebulae, namely nebulae where the confinement of the pulsar wind is provided by the ISM, in which the star supersonic motion produces a bow shock. 

The study of Bow Shock PWNe provides us with precious information about the particle release by such systems. First of all, we know from direct observations that the spectra of radio emitting particles, even in these older systems, are extremely similar to those observed in their younger counterparts\cite{ng,yous}: $N(E)\propto E^{-\gamma_e}$ with $\gamma_e\approx 1.5$. The high energy part of the spectrum is also similar to that observed in PWNe inside SNRs\cite{ng}. From the estimated value of the magnetic field we also know that radio emission is due to particles with energies between 1 GeV and 1 TeV,  and that these particles have a flat energy spectrum. In addition, numerical simulations\cite{bow05} support the idea that these low energy particles can easily escape from the system along the tail, with very little energy losses.

If we believe then that all the particles with energies up to 1TeV produced after a pulsar has left its parent SNR are released in the ISM with negligible losses, the main unknown we are left with in the quest for understanding whether pulsars can contribute appreciably to the positron excess, is $E_{\rm res}$, namely the fraction of the pulsar initial rotational energy that is still available for particle acceleration after a time $t_{\rm esc}$. This critical quantity depends on how the pulsar spins down. In general, one can write the spin-down law as $\dot \Omega \propto \Omega^n$, with $n$ the so-called braking index, and $\Omega$ and $\dot \Omega$ the star rotation frequency and its derivative respectively. For a spinning dipole one expects $n=3$, since $\dot E=I \Omega \dot \Omega \propto B_*^2 \Omega^4$, where $\dot E$ is the star's energy loss, $I$ its momentum of inertia and $B_*$ the surface magnetic field. However, the problem is that for the few cases in which the braking index has actually been measured, its value has always been found to be less than 3, with an average of 2.5. At the same time, while the surface magnetic field is typically found to be of order $a\ few \times 10^{12} G$, the distribution of initial spin periods is not well constrained. 

\begin{figure}
\includegraphics[scale=0.5]{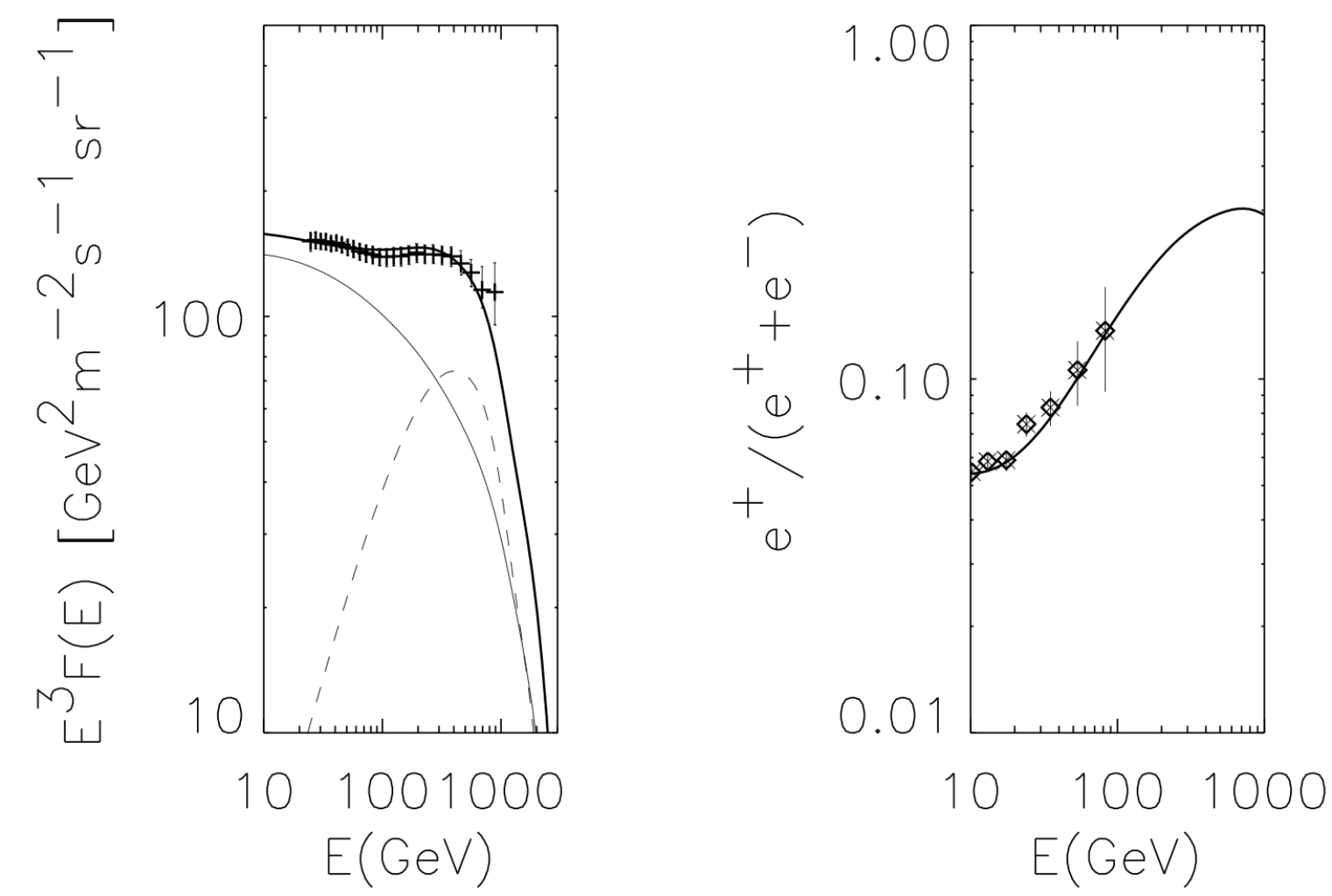}
\caption{Left panel: Spectrum of electrons from SNRs (solid line) and electrons and positrons from pulsar winds (dashed line). The thick solid line through the Fermi data points is the sum of the two. Right panel: Positron ratio compared with the PAMELA data points. Figure from Ref.~\protect\refcite{pospsr}\protect .}
\label{fig:elpos}
\end{figure}

Keeping in mind all these unknowns, it is still worth considering what comes out in terms of electrons and positrons from the simplest possible calculation\cite{pospsr}, namely one that considers all pulsars being born with a $10^{12} G$ surface magnetic field and a spin period of $20\ {\rm ms}$ (values close to those appropriate for the Crab pulsar), and leaving their parent remnant 40 kyr after birth. The spectrum of electrons and positrons at Earth is computed through a Green function approach, assuming that SNRs and pulsars are born at a rate of $1/(30\ yr)$ and are distributed in the Galaxy according to the results of Ref.~\protect\refcite{faucher}\protect. Fig.~\ref{fig:elpos} shows the results of such a calculation, both in terms of the total flux of leptons, which is compared with Fermi data in the left panel, and for the positron fraction, which is compared with PAMELA data on the right. This very simple model is clearly able to reproduce the data very well, assuming that after leaving the remnant our model pulsar releases about 20\% of its rotational energy in the form of pairs with energies between 1 GeV and 1 TeV.

An important probe of this model is in the level of predicted anisotropy: since, due to losses, the number of contributing sources, strongly decreases as a function of energy, with only a handful of sources contributing TeV leptons, an anisotropy at the 1\% level is predicted around this energy. Present day AMS-02 upper limits are at the 3\% level at an energy of 350 GeV. With more data, the level of anisotropy expected from models based on the pulsar origin of the excess positrons might be detectable.
  
\section{Conclusions}
\label{sec:concl}
In the last decade, there have been many important developments in the quest for understanding the origin of Galactic CRs. I have tried to review a few of them in this article. The most striking news have come from observations. Looking at SNRs with X-ray and $\gamma$-ray telescopes, we have seen the first unequivocal detection of relativistic protons and collected evidence for the presence of large magnetic fields, able to guarantee in principle acceleration up to the {\it  knee}. We might even have seen a remnant accelerating particles up to 1/2 PeV (Tycho), but some modelling is required to extract this information from the data and hence we cannot state this with absolute certainty, especially in the presence of the loose ends that are still there in the theory. Indeed the recent $\gamma$-ray data have also shown some discrepancies with respect to the theoretical expectations.

The last decade has been very active also on the theory side: the Non-Linear Theory of Diffusive Shock acceleration has become a fully developed framework for the description of a shock that is accelerating particles efficiently, with the progressive inclusion of all main sources of non-linearity of the problem, the back reaction of accelerated particles on the shock, the amplification of magnetic fields through instabilities induced by the accelerated particles and the back reaction of these fields on the system. The main difficulties that the theory currently faces are related to two pieces of evidence: 1) the observation in $\gamma$-ray emitting remnants of spectra that are systematically steeper than $E^{-2}$; 2) the conclusion reached, through a combination of observations and calculations of CR propagation in the Galaxy, that CR sources must inject in the Galaxy a spectrum steeper than $E^{-2}$. A promising way of accommodating these two pieces of evidence within the NLDSA theory is by redefinition of the velocity of the scattering centres with inclusion of the modified Alfv\'en speed, namely the Alfv\'en speed in the amplified magnetic field. Presently, this can only be done in a phenomenological way, in the absence of the necessary information on the properties of the relevant magnetic turbulence. However, the recent appearance in the literature of the first hybrid simulations of non-relativistic shock waves holds the promise of providing us with this information soon, together with new insights on another process that is currently treated in a phenomenological way, namely that of particle injection into the acceleration mechanism.  

In very recent times, the theory of NLDSA has been extended to shock propagating in partially ionised plasmas, where the non-thermal diagnostics can be supplemented by optical spectroscopy looking for anomalies in the Balmer line profiles that could be associated to efficient particle acceleration. Interestingly enough, it is exactly this method that has allowed us to find the best evidence so far of a SNR, RCW86, that is accelerating CRs with very high efficiency, of order 30\%.

In conclusion, in terms of proving the CR-SNR association, in the last few years we have really made important progress, collecting evidence of efficient acceleration of particles in SNRs, and seeing that these are not only electrons but also protons. One issue that cannot be considered as settled is that of the maximum achievable energy in SNRs, and the end of the Galactic CR spectrum. It is likely that the final word on the subject will have to wait the CTA era.

Aside from long standing questions, observational progress constantly provides us with new puzzles: two recent instances are the detection of spectral hardenings of basically all nuclear species at around 200 GeV/n and of a flatter spectrum of He with respect to protons. While the former finds a possible simple explanation in the effect of self-generated waves during CR propagation in the Galaxy, no simple explanation is currently available for the latter, which really stays as a big mystery. 

Finally, while we concluded that the SNR paradigm for the origin of CRs is in good shape, observations of the positron excess seem to require some extra source to intervene at energies well below 1 PeV to guarantee an extra input of relativistic leptons in the Galaxy. It is interesting to observe that if the pulsar origin of the excess discussed in this review will be confirmed, then all the sources so far needed of Galactic CRs would still be in the realm of stellar remnants.

\section*{Acknowledgments}
I am deeply grateful to Pasquale Blasi, for reading and commenting this manuscript, but even more for our long-term collaboration on this subject. I also want to thank Damiano Caprioli and Giovanni Morlino, with whom I have enjoyied collaborating on Cosmic Ray physics for several years.

\end{document}